\documentclass{article}
\usepackage{spconf,amsmath,graphicx}
\usepackage{booktabs}
\usepackage{multirow}
\usepackage{array}
\usepackage{url}
\usepackage{cite}
\usepackage{color, colortbl}
\usepackage{xcolor}
\colorlet{top5}{blue!10}
\colorlet{sig}{orange!10}
\colorlet{high_cor}{red!10}
\usepackage{subcaption}

\usepackage{makecell}

\title{The Singing Voice Conversion Challenge 2023}
\name{Wen-Chin Huang$^1$, Lester Phillip Violeta$^1$, Songxiang Liu$^2$, Jiatong Shi$^3$, Tomoki Toda$^1$}
\address{$^1$Nagoya University, Japan\\$^2$Tencent AI Lab\\$^3$Carnegie Mellon University, USA\\\texttt{svcc2023@vc-challenge.org}}

\begin{document}
\ninept
\maketitle
\begin{abstract}
We present the latest iteration of the voice conversion challenge (VCC) series, a bi-annual scientific event aiming to compare and understand different voice conversion (VC) systems based on a common dataset. This year we shifted our focus to singing voice conversion (SVC), thus named the challenge the Singing Voice Conversion Challenge (SVCC). A new database was constructed for two tasks, namely in-domain and cross-domain SVC. The challenge was run for two months, and in total we received 26 submissions, including 2 baselines. Through a large-scale crowd-sourced listening test, we observed that for both tasks, although human-level naturalness was achieved by the top system, no team was able to obtain a similarity score as high as the target speakers. Also, as expected, cross-domain SVC is harder than in-domain SVC, especially in the similarity aspect. We also investigated whether existing objective measurements were able to predict perceptual performance, and found that only few of them could reach a significant correlation.
\end{abstract}
\begin{keywords}
voice conversion, voice conversion challenge, singing voice conversion
\end{keywords}

\vspace{-3mm}
\section{Introduction}
\label{sec:intro}

Voice conversion (VC) refers to the task of converting one kind of speech to another without changing the linguistic contents \cite{vc-survey, vc-survey-2021}. VC has a wide range of applications covering from medical solutions to entertainment, such as speaking aid devices for patients \cite{EL-GMM, augmented-speech-production}, computer-assisted language learning leveraging accent conversion \cite{fac-for-call}, personalized expressive voice assistants \cite{personalized-expresive-tts} and silent speech interfaces \cite{silent-speech-interface}. It was believed that the underlying VC techniques are although shared but difficult to be compared, because of the various applications and the consequent datasets that are being used.

\begin{figure}[t]
	\centering
	\includegraphics[width=0.8\linewidth]{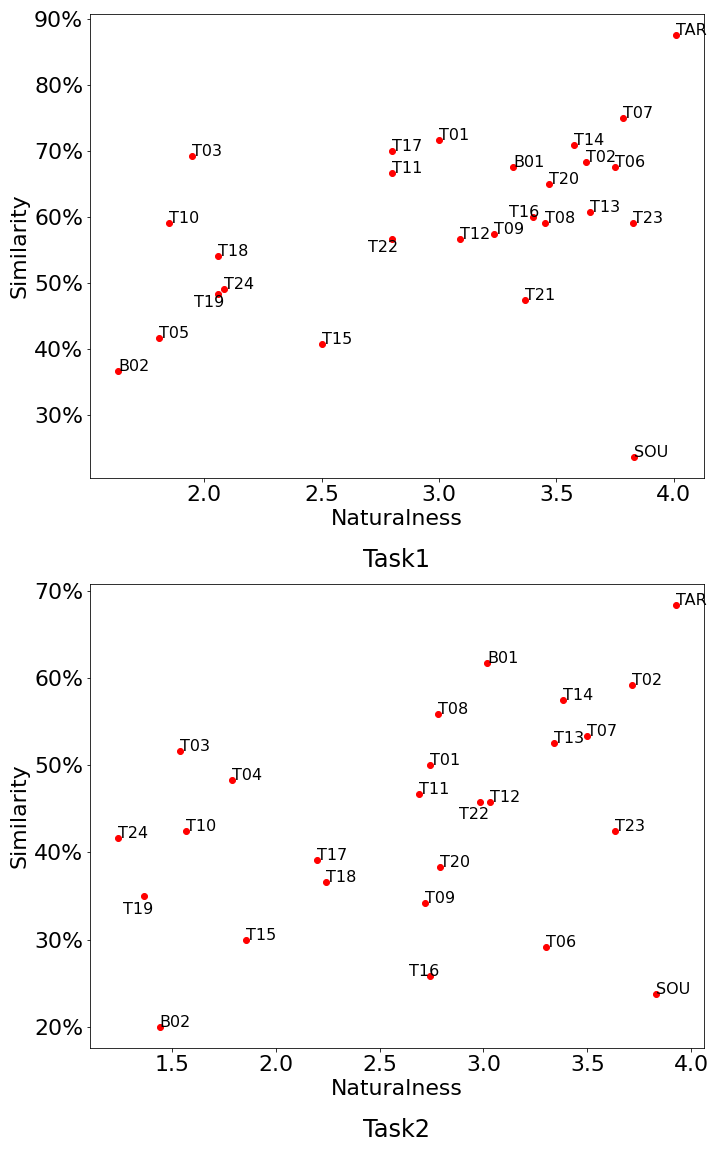}
    \vspace{-10pt}
	\caption{\label{fig:scatter}Scatter plots of naturalness and similarity percentage for task 1 (in-domain) and task 2 (cross-domain) from English listeners.}	
\vspace{-7mm}
\end{figure}

In light of this, the Voice Conversion Challenge (VCC) was first launched in 2016 \cite{vcc2016}, followed by there precedent versions in 2018 \cite{vcc2018} and 2020 \cite{vcc2020}. The objective of the VCC series was to better understand different VC techniques by looking at a common goal and dataset, and to share views about unsolved problems and challenges faced by current VC techniques. In the past three VCCs, speaker conversion, the transformation of speaker identity, which is long considered the most fundamental problem in VC, has been chosen as the main task. While the task remains unchanged, we gradually increased the difficulty, from parallel (supervised) training, non-parallel (unsupervised) training to cross-lingual conversion. In the latest challenge \cite{vcc2020}, it was shown that in terms of naturalness and speaker similarity, two important evaluation aspects in VC, the top system scored nearly as high as the ground truth of the target speakers. As described in Section~\ref{sec:related}, as VC techniques have significantly improved through the activities of these challenges, we decided to move on to a more challenging setting.

In 2023, we launched the fourth edition of the VCC, and by shifting our focus to singing voices, we renamed it the Singing Voice Conversion Challenge (SVCC). Singing voice conversion (SVC) is considered more challenging because: (1) compared to normal speech, it involves a wider range of varieties in pitch, energies, expressions, and singing style, (2) from the pitch information perspective, while the generated singing voice needs to follow the notes of the song, the singing style can vary from singer to singer, thus the level of disentanglement needs to be properly modeled. 
In the following sections, we describe the organization of the challenge and present the evaluation results of the submitted systems, where Figure~\ref{fig:scatter} shows a quick overview of the subjective results. 

\section{Related works}
\label{sec:related}
\subsection{Past voice conversion challenges}

\textbf{VCC2016} \cite{vcc2016} was held as a special session at INTERSPEECH 2016, and attracted 17 participants. A parallel VC database consisted of two source and two target native American English speakers (two females and two males), each of whom spoke 162 parallel sentences, was constructed for the only task in VCC2016. 
It was reported that the best system in VCC 2016 obtained an average naturalness score of 3.0 and a similarity score of 70\%\footnote{Defined as the percentage of a system's converted samples that were judged to be the same as the target speakers.}. However, it was obvious that there was a huge gap between the target natural speech and the converted speech.

\textbf{VCC2018} \cite{vcc2018} was held as a special session of the ISCA Speaker Odyssey Workshop 2018 and attracted 32 participants.
The two tasks were based on a newly constructed but smaller parallel VC database and a non-parallel VC database.
There were four native speakers of American English (two females and two males) for both the target and source speakers, each of whom uttered 80 sentences.
The same evaluation methodology in VCC 2016 was adopted for the 2018 challenge, and we observed significant progress. The best system performed well in both parallel and non-parallel tasks and obtained an average of 4.1 in naturalness and about 80\% in similarity. However, it was confirmed that there were still statistically significant differences between the target natural speech and the best converted speech in terms of both naturalness and speaker similarity.

\textbf{VCC2020} \cite{vcc2020} was held in a joint workshop with the Blizzard Challenge \cite{bc2020} and attracted 30 participants.
There was a semi-parallel intra-lingual conversion task and a cross-lingual conversion task, with two corresponding datasets newly built. For more details, please refer to \cite{vcc2020}.
The listening test results first showed that for the intra-lingual semi-parallel task, the speaker similarity scores of several systems were as high as the target speakers, while none of them achieved human-level naturalness. For the relatively harder cross-lingual task, although the overall naturalness and similarity scores were lower, the best systems had naturalness scores higher than 4.0 and similarity scores above  70\%.

\subsection{Singing voice conversion}
The task of SVC aims at converting the singing voice of a source singer to that of a target singer without changing the contents.
Mainstream SVC models can be categorized into two classes: 1) parallel spectral feature matching models and 2) information disentanglement based models. 
Early works on SVC use parametric statistical models, such as Gaussian mixture models (GMMs), to model source-target spectral conversion function leveraging parallel singing data \cite{gmmsvc1, gmmsvc2}. Parallel approaches based on generative adversarial networks (GANs) have also been proposed to improve conversion performance \cite{singan}.
Since parallel singing data is expensive to collect on a large scale, especially in multi-singer applications, researchers have investigated the use of non-parallel data for SVC. Both implicit and explicit information disentanglement methods have been studied to decompose voice identity, pitch, and linguistic content from a singing voice.
CycleGAN and StarGAN-based SVC models \cite{cycleganvc, StarGANv2VC} use adversarial training and cycle consistency loss to implicitly disentangle voice identity from other information including linguistic content, pitch information, etc.
The encoder-decoder framework is another hot topic in the research of SVC, which explicitly use either domain confusion loss or textual supervision to obtain pitch-invariant and singer-invariant content representation.
An auto-encoder-based unsupervised SVC model is studied, which uses speaker confusion techniques to disentangle singer information from the encoder output \cite{unsupervised-svc}. Based on this model, PitchNet \cite{pitchnet} employs an additional adversarial pitch confusion term to extract pitch-invariant and singer-invariant features from the encoder.
Rather than relying on domain confusion losses, various models separately train a content encoder model and an information-fusion decoder model to tackle the task of SVC. The encoder uses text supervision to obtain singer-invariant content features, either through phonetic posteriorgrams (PPGs) \cite{ppg-svc} or features extracted from some intermediate layer in an ASR acoustic model \cite{ucd-svc, fastsvc}. To increase the expressiveness of the model, it is likely that the decoder incorporates generative modeling, such as auto-regressive models \cite{ppg-svc}, GANs \cite{ucd-svc, fastsvc, zhou2022hifi}, or denoising diffusion probabilistic models (DDPMs) \cite{liu2021diffsvc}.

It is worthwhile to note that the SVC open-source community has been extremely active recently. The most popular project, \textit{so-vits-svc}, has over 15k stars on its Github repository\footnote{\url{https://github.com/svc-develop-team/so-vits-svc}, accessed on 2023.6.18.}. It is a collective effort of over 30 contributors, providing training scripts on a variety of encoders, acoustic models, and vocoders.

\section{Tasks, databases, and timeline for Singing Voice Conversion Challenge 2023}
Similar to the past VCC iterations, the primary objective is to conduct speaker conversion. For SVCC 2023, we separate the challenge into two any-to-one tasks: in-domain SVC and cross-domain SVC. The organizers developed a dedicated challenge dataset for the challenge and released the dataset in a manner that gave the participants around two months to train their models.

\noindent \textbf{Task 1: In-domain SVC}:
For Task 1, the main task was to convert to a target speaker, by using the target speakers' singing voices as training data. Compared to speech, the prosody of singing voices mostly follows musical notes rather than that of the spoken language. Although some previous VC methods could be directly applied to singing datasets, the main point of the task was to verify which methods could effectively replicate how the target singer sings the musical notes.

\noindent \textbf{Task 2: Cross-domain SVC}:
For Task 2, the main task was to convert to a target speaker, by using the target speakers' speech data. Compared to Task 1, Task 2 is generally considered harder as the model does not see how the target speaker's singing voice sounds, as the pitch range in the dataset is narrower compared to the one in Task 1. Moreover, a person's singing style cannot be seen from their speech alone. Although a more challenging task, it is important to note that Task 2 may be a more realistic and generalizable setting, as not all humans have the formal training to control their vocal cords and sing songs in the correct notes or key. 

\begin{table}[t]
\centering
    \footnotesize
    \caption{An overview of the SVCC 2023 dataset.}
    \label{tab:dataset}
    \vspace{-10pt}
\begin{tabular}{cccc}
\toprule
\textbf{SVCC 2023 ID} & \textbf{NHSS ID} & \textbf{Minutes} & \textbf{No. of phrases} \\
\midrule
IDM1 & M04 & 11.84 & 150   \\
IDF1 & F01 & 12.72 & 159  \\
CDM1 & M03 & 4.31 & 161  \\
CDF1 & F02 & 6.75 & 150 \\
\midrule
SM1 & M02 & 2.35 & 24  \\
SF1 & F04 & 2.39 & 24 \\
\bottomrule
\end{tabular}
\end{table}

\subsection{Dataset construction}
The SVCC 2023 database is a subset of the NUS-HLT Speak-Sing (NHSS) dataset \cite{nhss}. The original database is parallel in the sense that it contains a speaker's singing and speech data. Each speaker records 10 songs from a selection of 20 songs, making the dataset semi-parallel. For both tasks, we use six songs from each speaker as the training data. For the evaluation data, we used six phrases from each of the remaining four songs. We labeled the target singers for Task 1 with "ID" and the target speakers "CD" for Task 2. On the other hand, the source speakers were labeled with "S". Male speakers were labeled with "M", while female speakers were labeled with "F". An overview of the dataset's details is shown in Table \ref{tab:dataset}. An open-sourced script\footnote{\url{https://github.com/lesterphillip/SVCC23_FastSVC/tree/main/egs/generate_dataset}} can be used to generate the SVCC 2023 dataset from the NHSS dataset. Aside from the SVCC 2023 dataset, we allowed participants to use other external datasets, provided that these were open-sourced to allow reproducible experiments. 

\subsection{Timeline}
The challenge was first announced and promoted on January 19, 2023. Training data was released on February 17, 2023, while the evaluation data was released on April 21, 2023, giving participants around two months to develop their models. Participants were then asked to submit their converted results on April 28, 2023, along with a brief description of their systems.

\section{Participants and submitted systems}
\label{participants_systems}

\subsection{Challenge participants}

\begin{table}[t]
\scriptsize
\centering
\caption{\label{table:participants} {List of participant affiliations of SVCC 2023 in random order. In addition, five participants did not identify themselves.}}
\vspace{-10pt}
\begin{tabular}{|l|c|c|}
\hline
\multicolumn{1}{|c|}{\textbf{Affiliation}} & \textbf{Task 1} & \textbf{Task 2} \\ \hline \hline
University of Sheffield & Y & Y  \\ 
RIKEN Guardian Robot Project & Y & Y  \\ 
Duke Kunshan University & Y & N \\
WIZ.AI & Y & Y  \\ 
National Tsing Hua University & Y & N \\
Huya.Inc & Y & Y  \\ 
Advanced Micro Devices, Inc. & Y & Y  \\ 
Samsung Research China-Beijing & Y & Y  \\ 
\makecell[l]{TME Lyra Lab, Northwestern Polytechnical University,\\~~Xian Jiaotong University} & Y & Y  \\ 
Shanghai Jiao Tong University & Y & Y  \\ 
Bilibili Inc. & Y & Y  \\ 
The Chinese University of Hong Kong (Shenzhen) & Y & Y  \\ 
Northwestern Polytechnical University, TME Lyra Lab & Y & Y  \\ 
Soochow University & Y & Y  \\ 
Nagoya University & Y & Y  \\ 
Parakeet Inc. & Y & Y  \\ 
Federal university of Goiás (UFG) & Y & Y  \\ 
Individual 1 & Y & Y  \\ 
Individual 2 & Y & Y  \\
\hline
\end{tabular}
\vspace{-15pt}
\\ 
\end{table}

Table~\ref{table:participants} shows the participant affiliations and in which tasks they participated, listed in random order. In total, we have 24 submissions and 2 baselines systems, ending up with 25 and 24 systems in Tasks 1 and 2, respectively.
As in previous VCCs, we anonymized each team with a unique team ID (T01 to T24 for the participants and B01 and B02 for the baseline systems), and informed each team of their own team ID except for five participants who did not submit system descriptions despite repeated warnings from the organizers. The ordering is random and different from that in Table~\ref{table:participants}.

\subsection{Baseline systems}

\noindent \textbf{B01 (DiffSVC System)}:
The first baseline system is similar to the system presented in the DiffSVC paper \cite{liu2021diffsvc}, which was considered state-of-the-art as we organized this challenge. The detailed description is presented in Appendix \ref{appen:diffsvc}.

\noindent \textbf{B02 (Decomposed FastSVC System)}:
The second baseline system aims to provide a simple open-sourced baseline\footnote{\url{https://github.com/lesterphillip/SVCC23_FastSVC}} for the challenge. The network is similar to FastSVC \cite{fastsvc}, but decomposed into an acoustic model and a vocoder to reduce training time. A detailed description of the system is found in Appendix \ref{appen:B02}.

\subsection{Description of the submitted systems}

\begin{table}[t]
\caption{\label{table:taxonomy} {Details of participating systems in SVCC 2023.}}
\vspace{-10pt}
\footnotesize

\begin{tabular}{|c|l|l|l|} 
\hline
\textbf{ID} & \textbf{Content Feature} & \textbf{VAE} & \textbf{Vocoder} \\ 
\hline 
B01 & PPG & N & HiFi-GAN \\
B02 & HuBERT & N & HN-uSFGAN \\
\hline 
T01 & PPG & N & HiFi-GAN + BigVGAN$\ast$ \\
T02 & HuBERT & Y & DSPGAN \\
T03 & HuBERT & Y & HiFi-GAN \\
\hline 
T04 & \multicolumn{3}{l|}{Unknown}  \\
\hline 
T05 & PPG & N & HiFi-GAN \\
T06 & ContentVec & Y & N/A (nsf-HiFi-GAN)$\ddagger$ \\
T07 & HuBERT & Y & N/A (HiFi-GAN)$\ddagger$ \\
\hline 
T08 & \multicolumn{3}{l|}{Unknown}  \\
\hline 
T09 & Uncertain & Y & nsf-HiFi-GAN \\
T10 & WavLM & N & BigVGAN \\
T11 & PPG & N & HiFi-GAN \\
T12 & HuBERT & N & HiFi-GAN \\
T13 & ContentVec & N & SiFi-GAN \\
\hline 
T14 & \multicolumn{3}{l|}{Unknown}  \\
\hline 
T15 & None (Melspec)$\dagger$ & N & nsf-HiFi-GAN \\
T16 & PPG & N & BigVGAN \\
T17 & HuBERT & N & nsf-HiFi-GAN \\
\hline 
T18 & \multicolumn{3}{l|}{Unknown}  \\
\hline 
T19 & None (Melspec)$\dagger$ & N & HiFi-GAN \\
T20 & HuBERT & Y & nsf-HiFi-GAN \\
T21 & ContentVec & Y & nsf-HiFi-GAN \\
T22 & PPG+ContentVec & N & BigVGAN \\
T23 & PPG & Y & DSPGAN \\
\hline 
T24 & \multicolumn{3}{l|}{Unknown}  \\
\hline
\multicolumn{4}{l}{{\footnotesize"Unknown" implies teams who did not submit their system description.}} \\
\multicolumn{4}{l}{\footnotesize$\ast$: BigVGAN was used as a postfilter.} \\
\multicolumn{4}{l}{\footnotesize$\dagger$: No content feature as only the melspectrogram was used.} \\
\multicolumn{4}{l}{\footnotesize$\ddagger$: No vocoder was used since the decoder outputs waveform.}

\end{tabular}
\vspace{-15pt}
\\ 
\end{table}

\subsubsection{Common components}

Most systems this year adopt the recognition-synthesis (rec-syn) framework\footnote{Following the definition in \cite{s3prl-vc-jstsp}, any VC system that separately trains the recognizer and synthesizer can be categorized as the rec-syn framework.}, where several encoders (or recognizer) are first used to extract a set of features, including a content feature which contains compact linguistic or phonetic information from the input, and prosodic related features such as f0, energy, etc. Then, conversion is mostly carried out by a decoder (or synthesizer) to inject target information. The content feature encoder is usually trained to be speaker-independent, thus is assumed to be capable of handling any unseen speaker. The decoder training is often conducted by pre-training on a multi-speaker/singer dataset, and then fine-tuned on the target dataset or directly uses a speaker embedding to control the identity. Exceptions are T15 and T19, both of whom adopted StarGANv2-VC \cite{StarGANv2VC} which jointly trains the encoders and decoder. Finally, we noticed that most teams did not develop special techniques for individual tasks.

\subsubsection{Taxonomy}
\label{sssec:taxonomy}

While the VCC 2020 analysis paper \cite{vcc2020} analyzed the submitted systems by the feature conversion model and the vocoder, we found that the viewpoint should advance along with the development of VC techniques. This year, we base our analysis on three aspects that give the largest variations among different systems: content feature type, use of variational autoencoders (VAEs), and vocoder type. Note that the goal of this section is not to derive meaningful tendencies or scientific differences, but rather a trend of popular techniques used in the current moment.

\noindent{\textbf{Content feature type.}}
The content feature plays an important role in rec-syn based VC. A good content feature should be rich in content but contains little to no speaker information \cite{s3prl-vc-jstsp}. To facilitate this property, the PPG is a straightforward choice as it is derived from an ASR model which is trained in a supervised fashion to extract linguistic information. A total of 7 teams used PPGs. In recent years, self-supervised learning (SSL) based speech representations are drawing attention in VC as they benefit from large-scale unlabeled corpora and are shown to be able to disentangle speaker information. Among the 12 teams that used SSL speech representations, popular choices included HuBERT \cite{hubert}, WavLM \cite{wavlm} and ContentVec \cite{contentvec}.

\noindent{\textbf{Use of VAEs.}}
Introducing the VAE probabilistic framework in conditional generative models improves the generalization ability to unseen condition combinations \cite{portaspeech, vits}, which is essential in low-resource tasks like SVC. This is backed by the fact that, among the 8 teams that adopted VAE, many of them ranked in the top three in Tasks 1 and 2, as we will show in later sections.

\noindent{\textbf{Vocoder type.}}
Despite the development of end-to-end SVC systems \cite{unsupervised-svc}, we observed that most teams still adopt a two-stage framework such that a converted acoustic feature (mostly mel spectrogram) is first generated, and then a vocoder is used to generate the final waveform. Exceptions are T06 and T07, who directly trained their decoders to generate the converted waveform.
All vocoders used by this year's teams are still based on GANs, showing that these are still the most popular choice when it comes to vocoders, despite the progress in other generative frameworks like flow-based models or DDPMs. While 8 teams used the original HiFi-GAN \cite{hifigan}, 5 teams used its neural source filter (NSF) extension, which combines NSF \cite{nsf} to improve the generalization ability. Other popular choices include BiGVGAN \cite{bigvgan}, SiFi-GAN \cite{sifigan} and DSPGAN \cite{dspgan}. We noticed that both T23 and T02, the top systems for Tasks 1 and 2 in naturalness, respectively, adopted DSPGAN. However, this sample size is too small to conclude that DSPGAN is the ideal choice for SVC.

\noindent{\textbf{Other notable observations.}} Due to the scarcity of singing voice datasets, many teams included speech data to train their models. For example, T07 used more than 1000 hours of speech training data.
While 7 teams applied DDPMs, most teams still used classical deep learning frameworks like VAEs or GANs.
Finally, 5 teams mentioned that they directly based their system on the \textit{so-vits-svc} project.

\section{Subjective evaluation}

As in the previous VCCs, the perceptual study is considered the main evaluation method in SVCC 2023. Here we present, to our knowledge, by far the first large-scale subjective evaluation for SVC. In the following sections, we consider the results of the English subjects' main results.

\subsection{Listening test setup}

Two common aspects of VC are evaluated in this challenge, namely naturalness and similarity. The protocol was basically consistent with that in the previous VCCs, where listeners were asked to evaluate the naturalness on a five-point scale, and for conversion similarity, a natural target speech and a converted speech were presented, and listeners were asked to judge whether the two samples were produced by the same speaker on a four-point scale. For more details, please refer to the VCC 2020 paper \cite{vcc2020}.

Crowdsourcing on platforms like Amazon Mechanical Turk has been attractive in recent years thanks to its efficiency; however, it suffers from listener quality variations and trustworthy issues. Considering budget constraints, we followed the same protocol in VCC 2020 and outsourced the recruiting of listeners to two companies. Specifically, English and Japanese listeners were recruited by the Inter Group Corporation and Koto Ltd., respectively. The two sets of perceptual evaluation required a total of more than ¥700,000 Japanese yen.
Each evaluation set contained 53 webpages (25 systems for Task 1, 24 systems for Task 2, and source/target for Tasks 1 and 2), each of which contained one naturalness and one similarity question to evaluate the same sample. The numbers of total and average scores per system from the English/Japanese listeners are 12720/38160 and 120/360, respectively.

\subsection{Main results on English listeners}

\subsubsection{Naturalness}

\begin{figure}[t]
	\centering
	\includegraphics[width=0.9\linewidth]{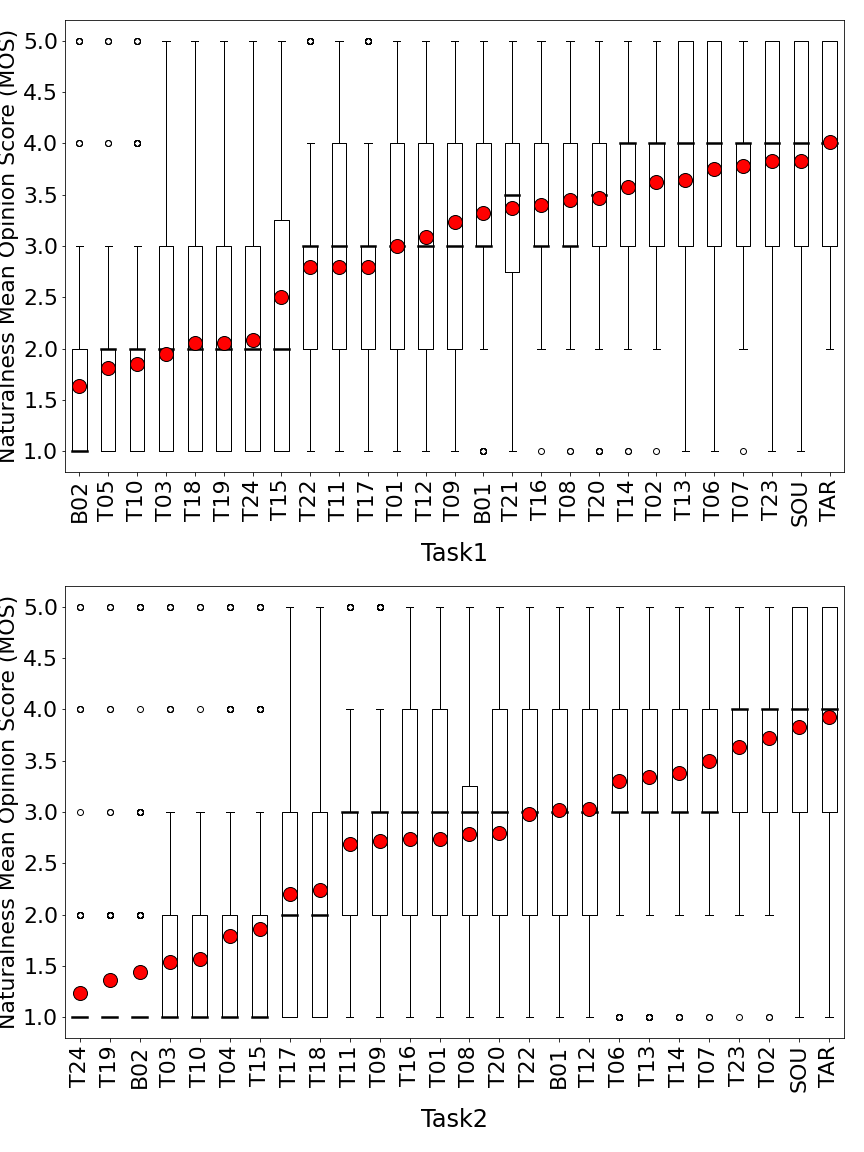}
	\caption{\label{fig:nat}Naturalness results for Tasks 1 and 2. MOS scores are arranged in accordance with their mean (red dot). SOU and TAR represents the source and target samples, respectively.
 }	
  \vspace{-15pt}
\end{figure}

\begin{figure}[t]
    \centering
    
    \begin{subfigure}[b]{0.49\columnwidth}
        \centering
        \includegraphics[width=\textwidth]{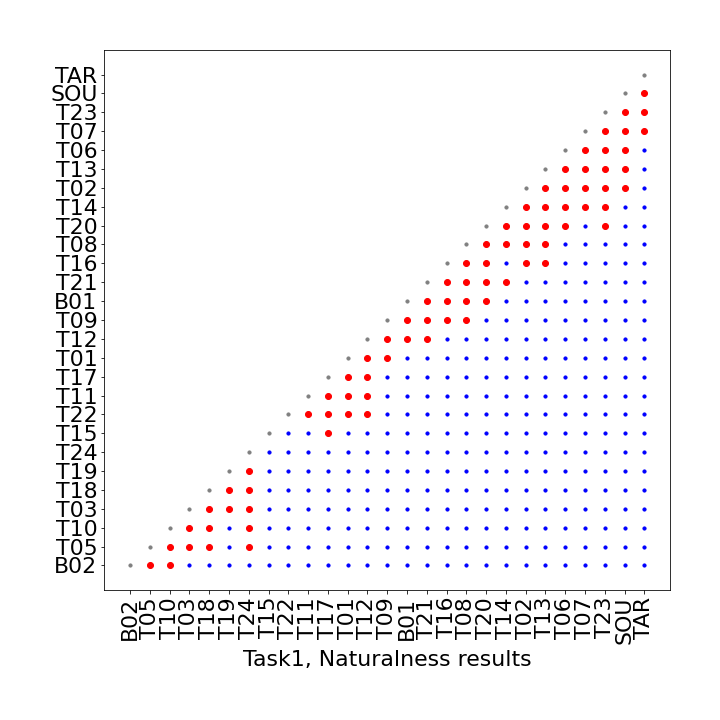}
        
        \caption{Task 1, naturalness}
        \label{fig:sig_en_task1_nat}
    \end{subfigure}
    \begin{subfigure}[b]{0.49\columnwidth}
        \centering
        \includegraphics[width=\textwidth]{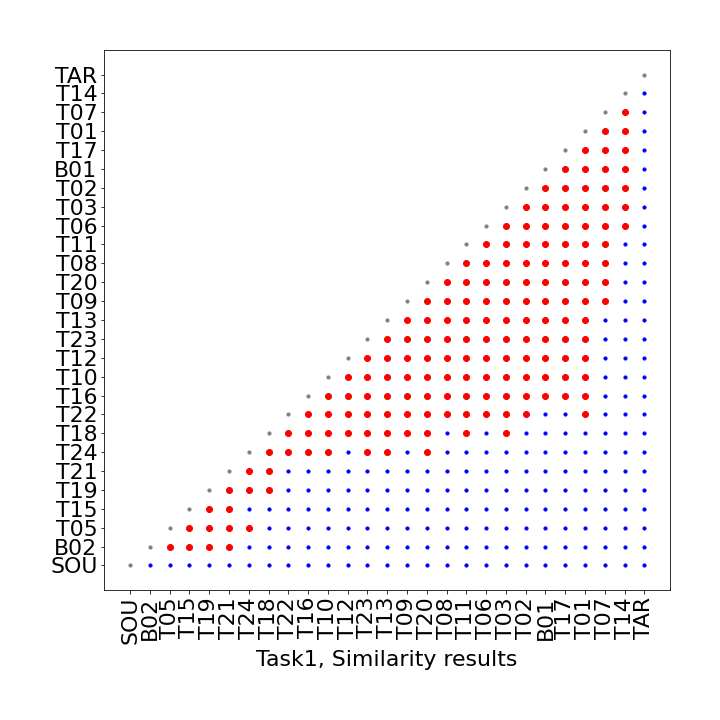}
        
        \caption{Task 1, similarity}
        \label{fig:sig_en_task1_sim}
    \end{subfigure}
    \\
    
    \begin{subfigure}[b]{0.49\columnwidth}
        \centering
        \includegraphics[width=\textwidth]{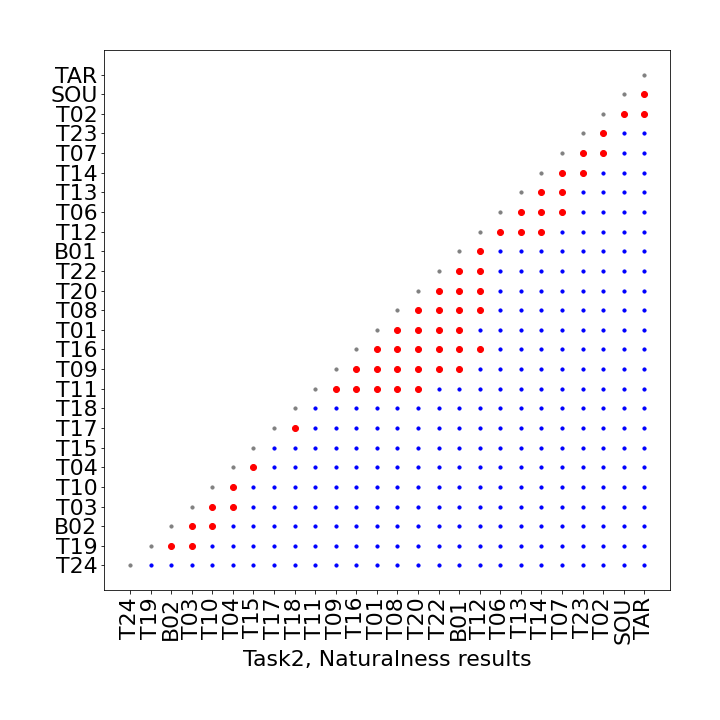}
        
        \caption{Task 2, naturalness}
        \label{fig:sig_en_task2_nat}
    \end{subfigure}
    \begin{subfigure}[b]{0.49\columnwidth}
        \centering
        \includegraphics[width=\textwidth]{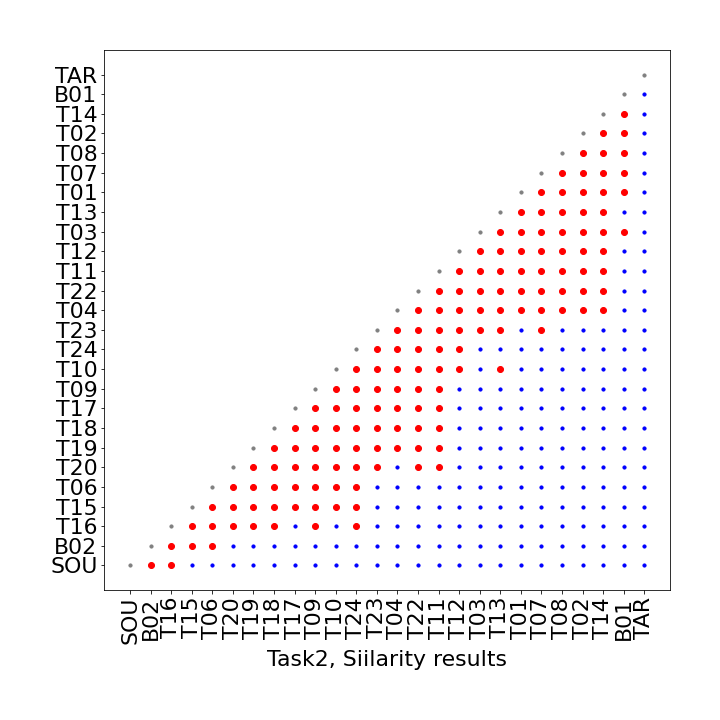}
        
        \caption{Task 2, similarity}
        \label{fig:sig_en_task2_sim}
    \end{subfigure}
    
    \centering
    \caption{Pairwise significance between systems, calculated with Wilcoxon signed-rank tests. Blue dots: significantly different; Red dots: no significant difference.}
    \vspace{-0.5cm}
    \label{fig:sig_en}
\end{figure}

Figure~\ref{fig:nat} shows the boxplot of the naturalness evaluation results of Tasks 1 and 2. First, baseline B01 was outperformed by around half and one-thirds of the teams in Tasks 1 and 2, respectively, showing that the SVC field has made significant progress in naturalness since DiffSVC was proposed. The top system in Task 1 was T23, which ranked second in Task 2. On the other hand, T02, the top system in Task 2, ranked fifth in Task 1. Although no system had a mean score higher than those of the source and target, Figures~\ref{fig:sig_en_task1_nat} and~\ref{fig:sig_en_task2_nat} show that T23, T07, and T02 are in fact not significantly different from the natural samples, showing that \textbf{the top systems have reached human-level naturalness}. Finally, we can see that in Task 2, only 8 teams scored more than 3.0, compared to the 14 teams in Task 1, showing that cross-domain SVC is indeed harder than in-domain SVC.

\begin{figure}[t]
	\centering
	\includegraphics[width=0.9\linewidth]{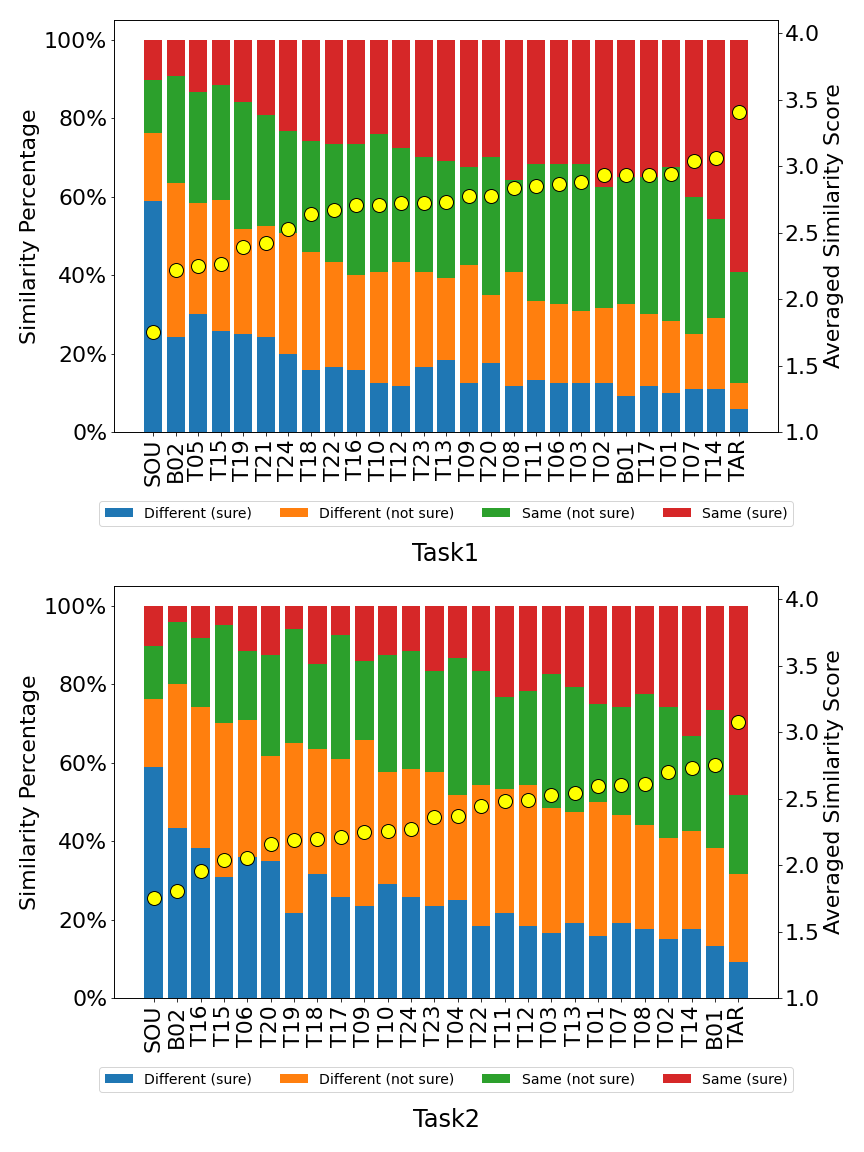}
	\caption{\label{fig:sim}Similarity results for Tasks 1 and 2. Similarity scores are arranged in accordance with their mean value (red dot). SOU and TAR represents the source and target samples, respectively.}
 \vspace{-15pt}
\end{figure}

\subsubsection{Similarity}

Figure~\ref{fig:sim} shows the results for the similarity evaluation results of Tasks 1 and 2. The similarity percentage is defined as the sum of the percentages from the ``same (not sure)'' and ``same (sure)'' categories, and the averaged scores are also shown. First, different from the naturalness results, baseline B01 ranked the fifth and the first in Tasks 1 and 2, respectively. T14, the top system in Task 1, ranked the second in Task 2. In contrast to naturalness, there is a clear gap (around 0.4 points) between the target samples and the top system in both tasks. This can also be observed from Figures~\ref{fig:sig_en_task1_sim} and~\ref{fig:sig_en_task2_sim}, which show that the target samples were significantly better than all other systems in terms of similarity. In conclusion, \textbf{when it comes to similarity in SVC, there is still a large room for improvement}. 

Similar to naturalness, there is a significant similarity degradation in Task 2. To our surprise, even the target samples suffer from such a drop (3.4 v.s. 3.0). As we manually inspected the natural samples, we found that due to the high variation of singing voices, different phrases of the same singer in the same song can sound like different people. We hypothesize that this makes the evaluation more difficult, as we observe that, from Figures~\ref{fig:sig_en_task1_sim} and~\ref{fig:sig_en_task2_sim}, when it comes to similarity, it is harder for listeners to distinguish between different teams (more red dots compared to Figures~\ref{fig:sig_en_task1_nat} and~\ref{fig:sig_en_task2_nat}).

Figure~\ref{fig:scatter} shows the scatter plots of naturalness and similarity percentage of both tasks. It can be clearly observed that there is a trade-off between naturalness and similarity for most systems, i.e. no team is dominant in both naturalness and similarity. This implies that all teams need to improve either similarity or naturalness.

\subsection{Do Japanese listeners make similar judgments compared with English listeners?}
\label{ssec:enjp}

We investigate whether non-native listeners (Japanese listeners in our case) perceive naturalness and similarity in SVC differently compared to English listeners. First, the linear correlation coefficients of the scores from English and Japanese listeners are 0.985, 0.975, 0.985 and 0.924 in Task 1 naturalness, Task 1 similarity, Task 2 naturalness and Task 2 similarity, respectively. Despite the high correlation, to examine whether there exists biases between English and Japanese listeners, we show their scatter plots in Figure~\ref{fig:enjp}. In general, we found that Japanese listeners tend to give higher scores in naturalness, and English listeners tend to give higher scores in similarity. Consequently, Japanese listeners can hardly distinguish natural singing voices from the converted ones by the top systems. Our second observation is based on Figures~\ref{fig:sig_en} and~\ref{fig:sig_jp}. Since the total number of scores received from English listeners is smaller, it is expected that it will be harder for them to distinguish between different systems (that is, more red dots should be observed). While this hypothesis somehow stands for similarity, it is surprising to see that for naturalness, English listeners can reach a similar level of confidence with only one-third of the scores. This result somehow implies that native listeners are more confident in evaluating naturalness.

\section{Objective evaluation}

\begin{table*}[ht]
\centering

\caption{\label{tab: spearman}Spearman correlation between objective and subjective metrics. Highlights in red indicate the highest correlation with corresponding subjective metrics among the objective metrics. CER metric refers to Conformer-based speech recognition results, while CER+ refers to Whisper results. Significance levels are shown by * (Significance levels: ***$p<$0.01, **$p<$0.05, *$p<$0.1).}
\vspace{-10pt}
\begin{tabular}{l|l|llllllllll}
\toprule
\textbf{Sub. Score} & \textbf{Listener} & \textbf{MCD} & \textbf{F0RMSE} & \textbf{F0CORR} & \textbf{CER} & \textbf{CER+} & \textbf{$D_{\text{Embed}}$} & \textbf{UTMOS} & \textbf{SSL-MOS}   \\
\midrule
\multirow{2}{*}{Task 1 MOS} & JPN & -0.28 & -0.41** & 0.48** & -0.62*** & \cellcolor{high_cor}-0.80*** & -0.58*** & 0.77*** & 0.53*** \\
 & ENG  & -0.24 & -0.28 & 0.45** & -0.57*** & \cellcolor{high_cor}-0.73*** & -0.45** & 0.72*** & 0.42 \\
  \midrule
\multirow{2}{*}{Task 1 SIM} & JPN & -0.62*** & -0.26 & 0.37* & -0.42** & -0.40** & \cellcolor{high_cor}-0.83*** & 0.49** & 0.30 \\
 & ENG  & -0.45** & -0.10 & 0.21 & -0.26 & -0.27 & \cellcolor{high_cor}-0.63*** & 0.38* & 0.13 \\
  \midrule
\multirow{2}{*}{Task 2 MOS} & JPN & -0.38* & -0.27 & 0.10 & -0.62*** & \cellcolor{high_cor}-0.77*** & -0.58*** & 0.60*** & 0.15 \\
 & ENG  & -0.29 & -0.06 & -0.16 & -0.60*** & \cellcolor{high_cor}-0.73*** & -0.45** & 0.49** & 0.11 \\
  \midrule
\multirow{2}{*}{Task 2 SIM} & JPN & -0.38* & \cellcolor{high_cor}-0.67*** & 0.03 & -0.25 & -0.53*** & \cellcolor{high_cor}-0.67*** & -0.08 & -0.27 \\
 & ENG  & -0.29 & -0.22 & -0.37* & -0.11 & -0.28 & \cellcolor{high_cor}-0.41** & -0.20 & -0.23 \\
\bottomrule
\end{tabular}
\vspace{-15pt}
\end{table*}

\subsection{Objective metrics}

Similar to the previous VCCs, we investigate objective evaluation metrics for SVC submissions to motivate future evaluation of SVC research. Specifically for this year, we adopt objective metrics focusing on spectrogram distortion, F0, intelligibility, singer similarity, and neural predictors for naturalness. 

\noindent{\textbf{Spectrogram distortion}}: We use mel cepstral distortion (MCD) for the evaluation of spectrogram distortion following previous works \cite{singan, villavicencio2010applying, gmmsvc1, doi2012singing, liu2021diffsvc, lu2020vaw, shi2022muskits}. The implementation follows \cite{huang2020voice, huang2021pretraining}.

\noindent{\textbf{F0 metrics}}: F0-related metrics have been widely used in previous SVC works \cite{liu2021diffsvc, ucd-svc, pitchnet, rajpura2020effectiveness, ppg-svc, gmmsvc2, zhou2022hifi}. For this challenge, we select F0 Root Mean Square Error (RMSE) and correlation coefficient (CORR) as our objective metrics.

\noindent{\textbf{Intelligibility}}: Lyrics are an important component of singing voices. Previous investigations in voice conversion challenges \cite{vcc2020} have shown that speech recognition error rate could be an essential indicator of the system's performance. In this work, we utilize two Automatic Speech Recognition (ASR) models to conduct lyrics recognition and use the Character Error Rate (CER) as the evaluation metric. Specifically, we adopt a pre-trained HuBERT-based Conformer-based model trained over \texttt{dsing} corpus\footnote{\url{https://huggingface.co/espnet/ftshijt_espnet2_asr_dsing_hubert_conformer}} \cite{Roa_Dabike-Barker_2019} with ESPnet+S3PRL \cite{watanabe2018espnet, guo2021recent, yang21c_interspeech} and the whisper-large ASR model\footnote{\url{https://github.com/openai/whisper}} \cite{radford2022robust}.

\noindent{\textbf{\textbf{Speaker similarity}}: Previous works in SVC have explored using a singer identification/verification model to estimate singer conversion accuracy \cite{unsupervised-svc, ucd-svc, luo2020singing}. Some other works also estimate the singer similarity with pre-trained singer embedding \cite{zhou2022hifi, fastsvc}. In this work, we adopt RawNet-3 based speaker embedding \cite{jung2022pushing} to estimate the singer similarity by calculating their cosine similarity (i.e., $D_{\text{embed}}$).

\noindent{\textbf{\textbf{Neural MOS predictor}}: As a previous work also use Neural MOS predictor \cite{zhou2022hifi}, we also examine the performance of pretrained neural MOS predictor with the baseline system (SSL-MOS) and best system (UTMOS) in VoiceMOS Challenge 2022 \cite{cooper2022generalization, saeki22c_interspeech}.

\subsection{Analysis with subjective evaluation}

In order to examine the relationship between subjective and objective evaluation metrics, we computed the Spearman correlation coefficients for each metric. The detailed results can be found in Table~\ref{tab: spearman}. (1) In most cases, metrics related to spectrogram and fundamental frequencies do not exhibit a significant correlation with subjective evaluation, which diverges from the findings of previous studies on VC in speech. (2) Speech recognition measures, both for the Conformer-based recognizer and Whisper, demonstrate a noteworthy correlation with the subjective MOS. (3) Currently, it is challenging to accurately assess singer similarity using objective metrics. The singer embedding cosine distance performs the best among the metrics, showing statistical significance for both Task 1 and Task 2 evaluations among Japanese and English speakers. However, even this metric yields insignificant results when assessing the similarity of Task 2 subjective measures with native speakers. (4)~Despite being trained on speech corpora, the existing MOS predictor, UTMOS, exhibits a moderate correlation with subjective measures of naturalness, indicating its generalization capability.

\section{Conclusion}

The singing voice conversion challenge 2023 is the fourth edition of the voice conversion challenge series, held to compare and understand different VC systems built on a common dataset. We introduced two tasks, namely any-to-one in-domain SVC and any-to-one cross-domain SVC, and curated a database which is essentially a subset of the NHSS dataset.
After giving participants two and a half months to train their SVC systems, we received a total of 26 submissions, including 2 baselines. As the first large-scale listening test for SVC, we observed that the top SVC systems in both tasks have achieved human-level naturalness. However, we also confirmed that there is a significantly large gap between the similarity scores of the target and all submitted systems. In addition, we confirmed that the cross-domain task is indeed a more difficult task, as the overall naturalness and similarity scores were lower. Finally, we showed that as few objective evaluation metrics can moderately correlate with the subjective scores, even the metric that best correlates with the similarity scores only yields a weak correlation, showing that objective assessment for SVC still has a lot to improve.

\section{ACKNOWLEDGMENTS}
\label{sec:ack}

This work was partly supported by JSPS KAKENHI Grant Number 21J20920, and JST CREST Grant Number JPMJCR19A3. We thank Yusuke Yasuda from Nagoya University for his assistance in conducting the listening test.

\bibliographystyle{IEEEbib}
\bibliography{refs}

\clearpage
\appendix
\section{Listener Details}
\label{appen:listener}

We recruited 40 unique English listeners (17 female, 22
male, and 1 unknown), and Figure~\ref{fig:accent_age} shows the accent and age distributions of the English and Japanese listeners. Half of the English participants were in their 30s or 40s, and most of them had an American accent. 
For Japanese listeners, we had a total of 319 unique valid listeners
(162 male and 157 female). Figure~\ref{fig:accent_age} also shows that most of the Japanese listeners were in their 30s or 40s.

\begin{figure}[h]
	\centering
	\includegraphics[width=0.5\linewidth]{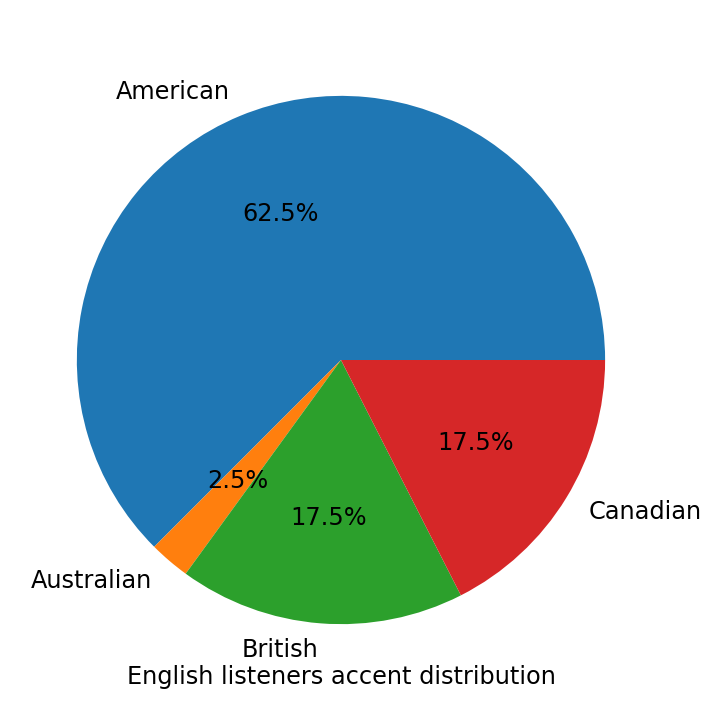} \\
	\includegraphics[width=0.5\linewidth]{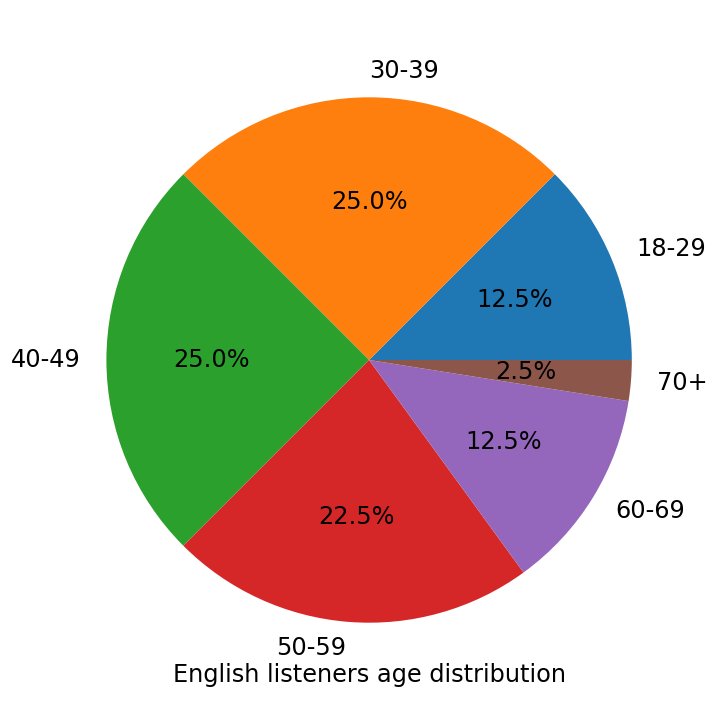} \\
        \includegraphics[width=0.5\linewidth]{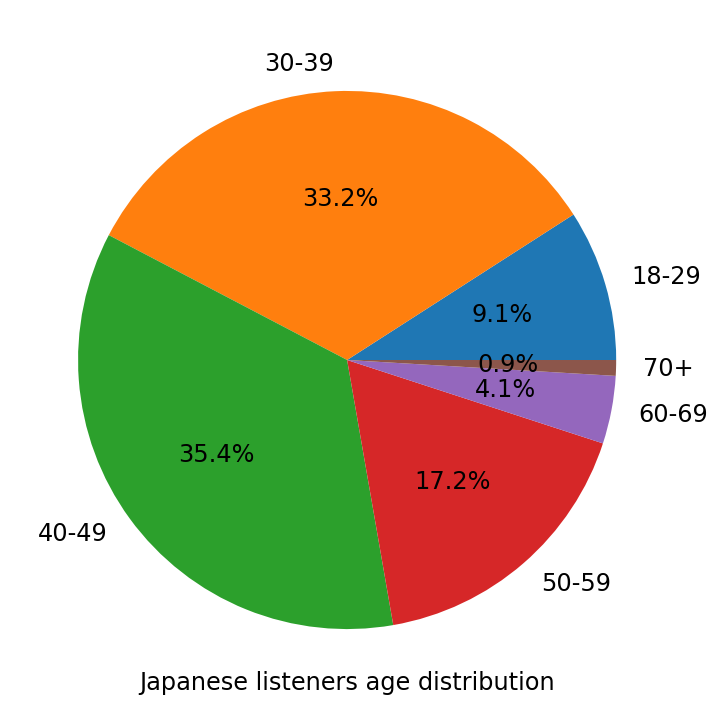} \\
	\caption{Age and accent distribution of English and Japanese listeners.	\label{fig:accent_age}}	
\end{figure}

\section{Evaluation results from Japanese listeners}
\label{appen:results_jp}

\subsection{Naturalness}

Figure~\ref{fig:nat_jp} shows the boxplot of the naturalness evaluation results of Tasks 1 and 2 from the Japanese listeners. In both tasks, baseline B01 was outperformed by around half of the teams. T23 was the top system in both tasks. Surprisingly, different from the finding from the English listener results that no team was on average better than the natural samples (TAR, SOU), three teams (T23, T07, T02) and one team (T23) received a naturalness score higher than the natural samples in Tasks 1 and 2, respectively. Furthermore, the pairwise significance test results in Figures~\ref{fig:sig_jp_task1_nat} and~\ref{fig:sig_jp_task2_nat} show that the natural samples are not significantly different with six teams (T23, T07, T02, T06, T14, T20) and two teams (T23, T02) in Tasks 1 and 2, respectively. These findings are in line with our finding that Japanese listeners tend to give higher scores than those given by English listeners. Finally, we can also observe that the scores received in Task 2 are generally lower than those received in Task 1, again showing that cross-domain SVC is indeed harder than in-domain SVC.

\begin{figure}[h]
	\centering
	\includegraphics[width=\linewidth]{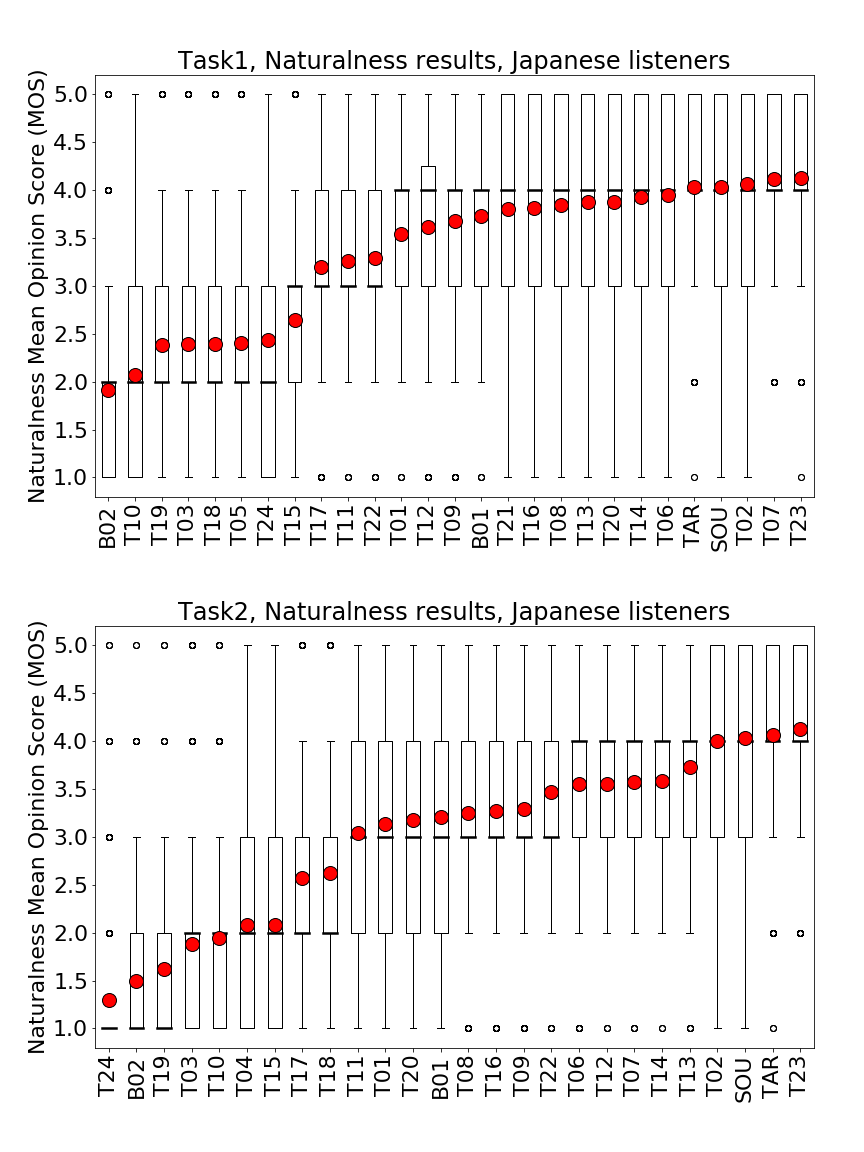}
	\caption{\label{fig:nat_jp}Japanese listeners' naturalness results for Tasks 1 and 2. MOS scores are arranged in accordance with their mean (red dot). SOU and TAR represent the source and target samples, respectively.
 }	
\end{figure}

\subsection{Similarity}

Figure~\ref{fig:sim_jp} shows the results for the similarity evaluation results of Tasks 1 and 2 from the Japanese listeners. Again, the similarity percentage is defined as the
sum of the percentages from the ``same (not sure)'' and ``same (sure)'' categories, and the averaged scores are also shown. The baseline B01 received a stronger ranking from Japanese listeners, ranking second in both tasks. The top system in Task 1, T14, ranked third in Task 2, while the top system in Task 2, T02, ranked fourth in Task 1. Similar to English listeners' results, there is also a clear gap (around 0.4 points) between the target samples and the top system in both tasks. This can also be observed from Figures~\ref{fig:sig_jp_task1_sim} and~\ref{fig:sig_jp_task2_sim}, which show that the target samples were significantly better than all other systems in terms of similarity. The conclusion is therefore similar to that of the English listeners: there is still a lot to work on for similarity.

\begin{figure}[h]
	\centering
	\includegraphics[width=\linewidth]{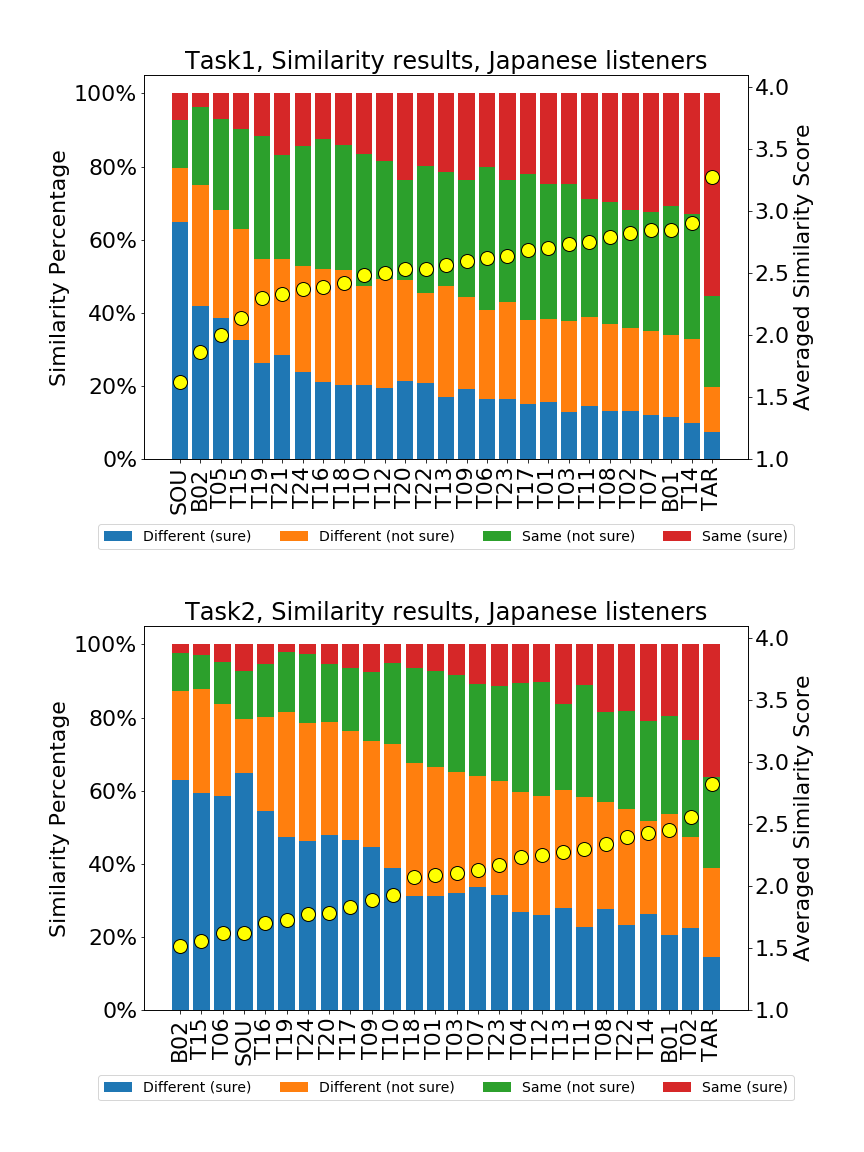}
	\caption{\label{fig:sim_jp}Japanese listeners similarity results for Tasks 1 and 2. Similarity scores are arranged in accordance with their mean value (red dot). SOU and TAR represents the source and target samples, respectively.}
\end{figure}

Figure~\ref{fig:scatter_jp} shows the scatter plots of naturalness and similarity percentage of both tasks from Japanese listeners. Similar to English listeners results, there is a trade-off between naturalness and similarity for most systems, i.e. no team is dominant in both naturalness and similarity. Again, all teams need to improve either similarity or naturalness.

\begin{figure}[h]
	\centering
	\includegraphics[width=\linewidth]{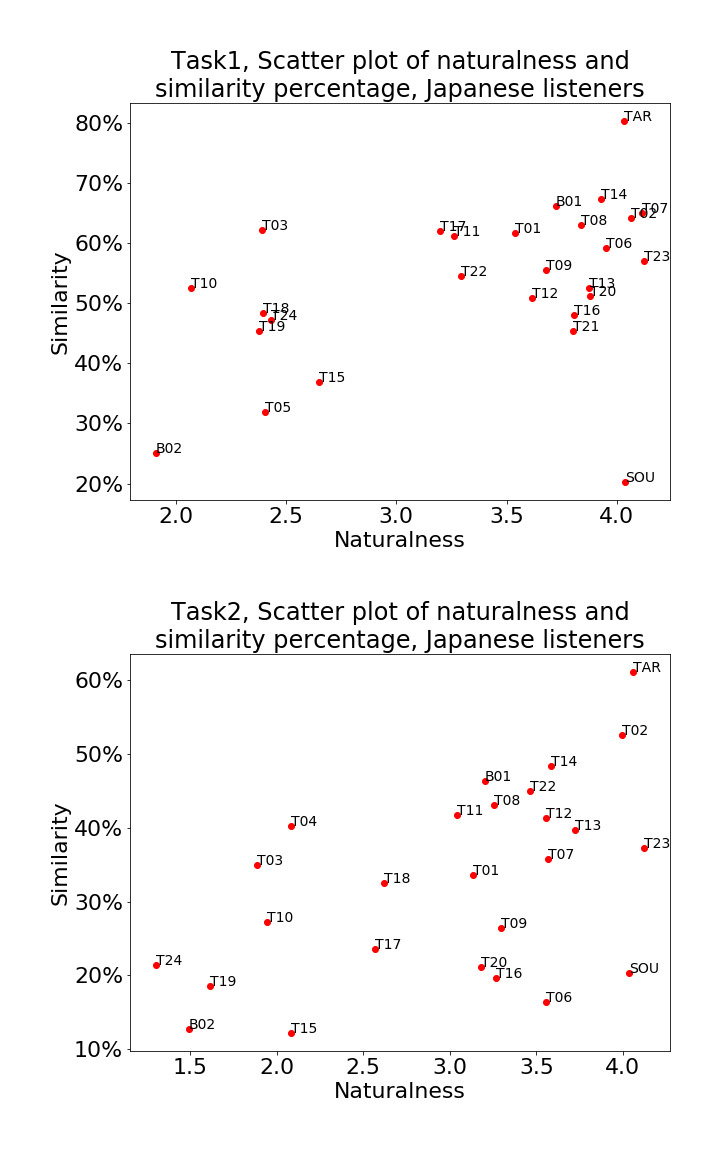}
	\caption{\label{fig:scatter_jp}Scatter plots of naturalness and similarity percentage for task 1 (in-domain) and task 2 (cross-domain), from Japanese listeners.}	
\end{figure}

\begin{figure}[h]
    \centering
    
    \begin{subfigure}[b]{0.49\columnwidth}
        \centering
        \includegraphics[width=\textwidth]{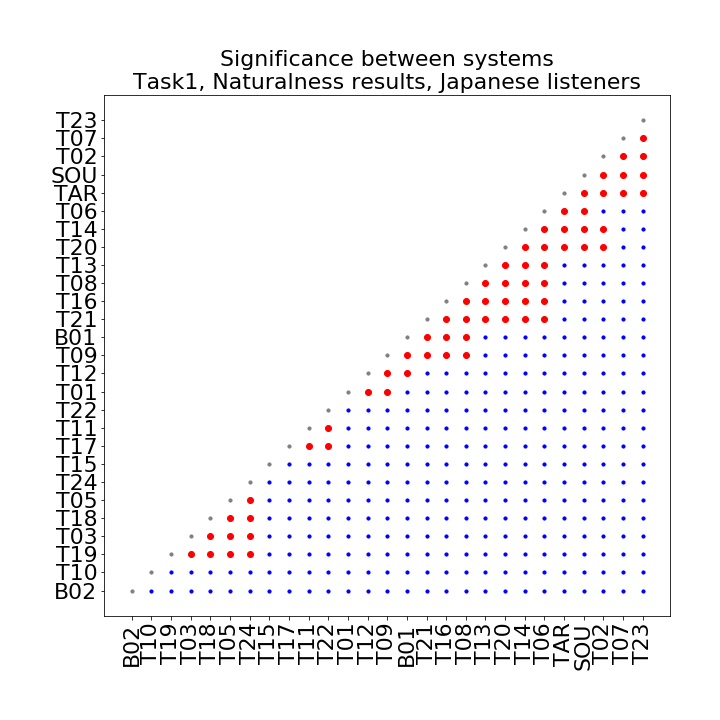}
        
        \caption{Task 1, naturalness}
        \label{fig:sig_jp_task1_nat}
    \end{subfigure}
    \begin{subfigure}[b]{0.49\columnwidth}
        \centering
        \includegraphics[width=\textwidth]{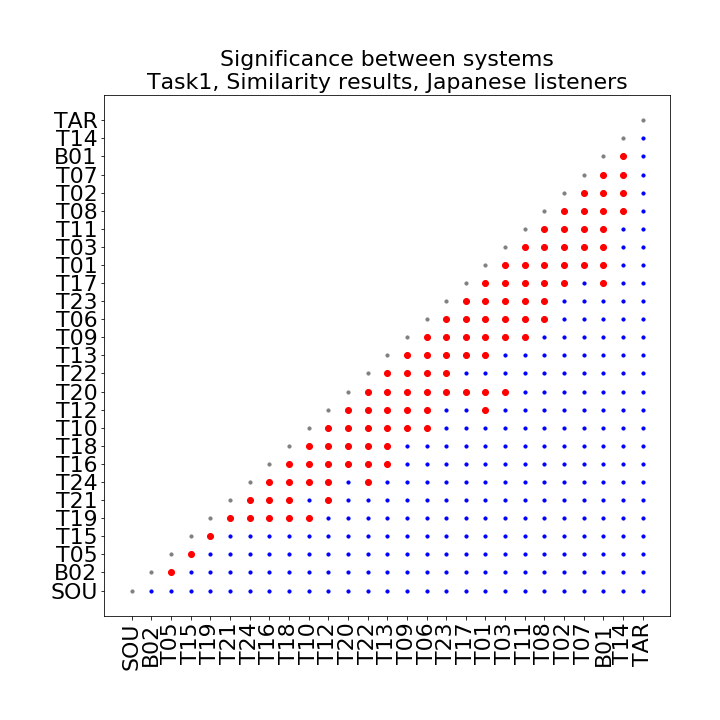}
        
        \caption{Task 1, similarity}
        \label{fig:sig_jp_task1_sim}
    \end{subfigure}
    \\
    
    \begin{subfigure}[b]{0.49\columnwidth}
        \centering
        \includegraphics[width=\textwidth]{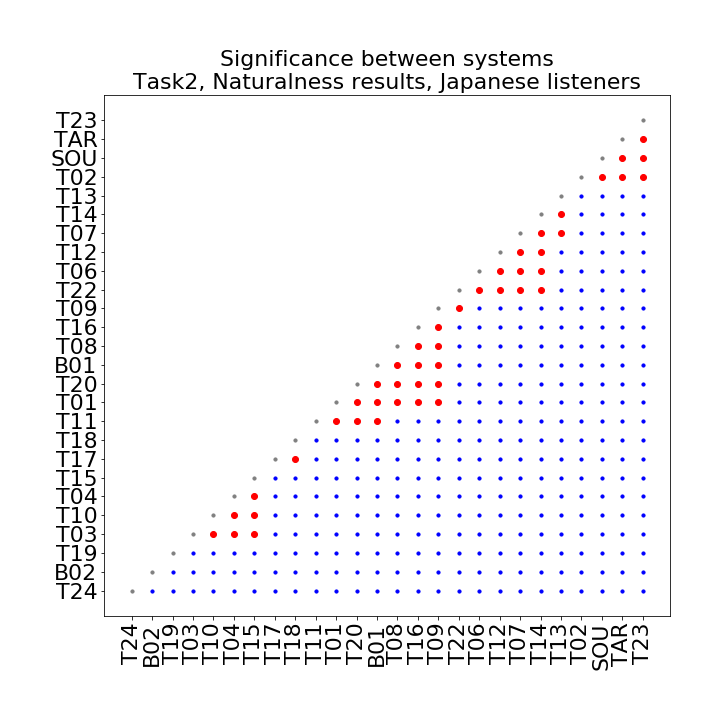}
        
        \caption{Task 2, naturalness}
        \label{fig:sig_jp_task2_nat}
    \end{subfigure}
    \begin{subfigure}[b]{0.49\columnwidth}
        \centering
        \includegraphics[width=\textwidth]{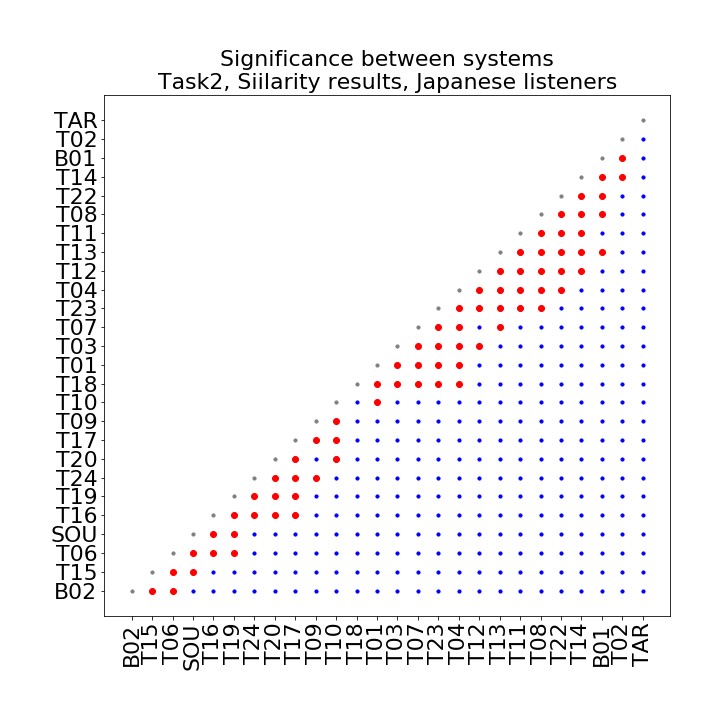}
        
        \caption{Task 2, similarity}
        \label{fig:sig_jp_task2_sim}
    \end{subfigure}
    
    \centering
    \caption{Japanese listeners pairwise significance between systems, calculated with Wilcoxon signed-rank tests. Blue dots: significantly different; Red dots: no significant difference.}
    \vspace{-0.3cm}
    \label{fig:sig_jp}
\end{figure}

\section{Comparison between English and Japanese listeners}
\label{appen:enjp}

Figure~\ref{fig:enjp} shows the scatter plots from Japanese and English listeners, and it can be observed that Japanese listeners tend to give higher scores in naturalness, and English listeners tend to give higher scores in similarity.

We made a hypothesis in Section~\ref{ssec:enjp} that the larger the number of scores, the easier it is to observe statistically significant differences between systems, which means fewer red dots should be observed in Figures Figure~\ref{fig:sig_jp} and~\ref{fig:sig_en}. However, by comparing Figure~\ref{fig:sig_jp} (Japanese listeners) and Figure~\ref{fig:sig_en} (English listeners), we observed that while this hypothesis somehow stands for similarity, it is surprising to see that for naturalness, English listeners can reach a similar level of confidence with only one-thirds of scores.

\begin{figure}[t]
    \centering
    
    \begin{subfigure}[b]{\columnwidth}
        \centering
        \includegraphics[width=\textwidth]{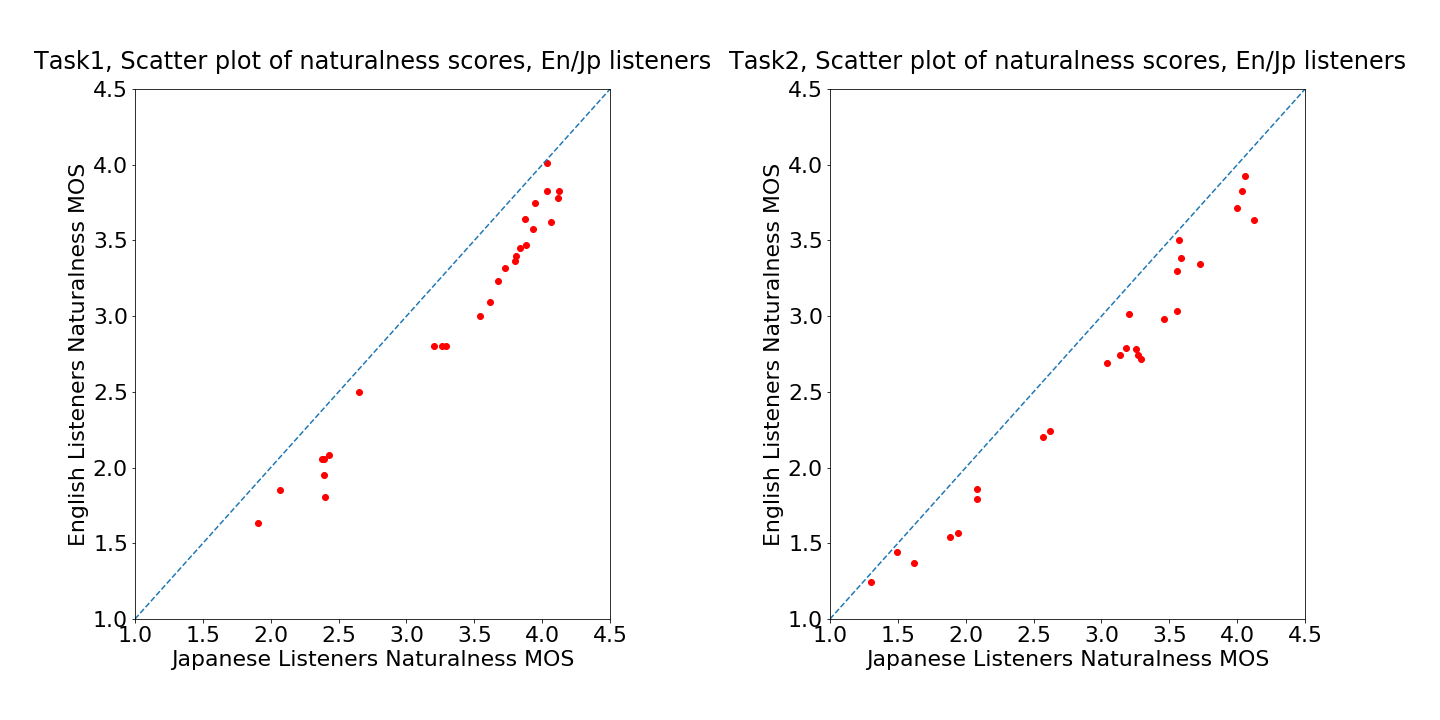}
        \caption{Naturalness}
        \label{fig:nat_enjp}
    \end{subfigure}
    \\
    
    \begin{subfigure}[b]{\columnwidth}
        \centering
        \includegraphics[width=\textwidth]{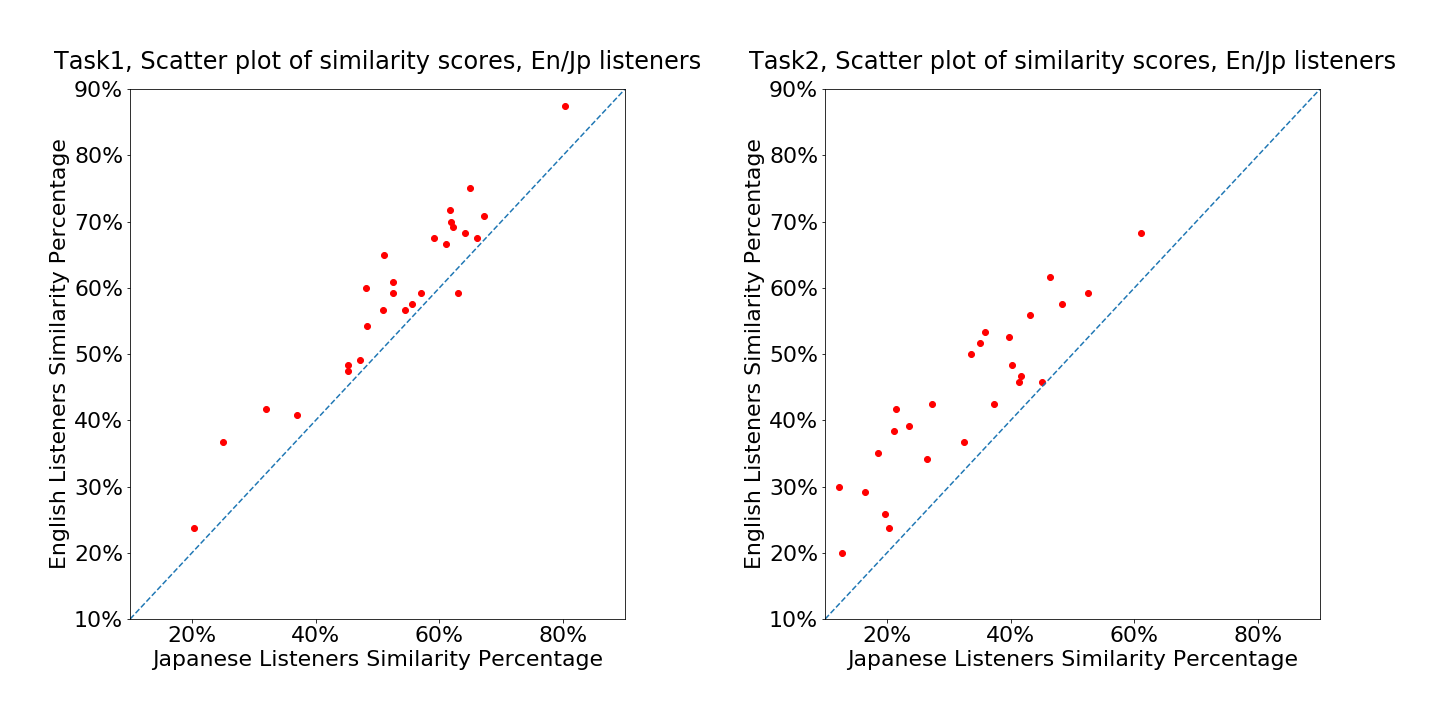}
        
        \caption{Similarity}
        \label{fig:sim_enjp}
    \end{subfigure}
    
    \centering
    \caption{Scatter plots of scores from Japanese listeners and English listeners.}
    \vspace{-0.3cm}
    \label{fig:enjp}
\end{figure}

We further examine whether the above-mentioned hypothesis implies the following statement: the larger the number of scores, the smaller the system-level variance. We therefore plotted the system-level variance from Japanese and English listeners in Figure~\ref{fig:var_enjp}. However, we did not observe any obvious tendency, thus the above-mentioned statement was not implied in this challenge. 

\begin{figure}[h]
	\centering
	\includegraphics[width=\linewidth]{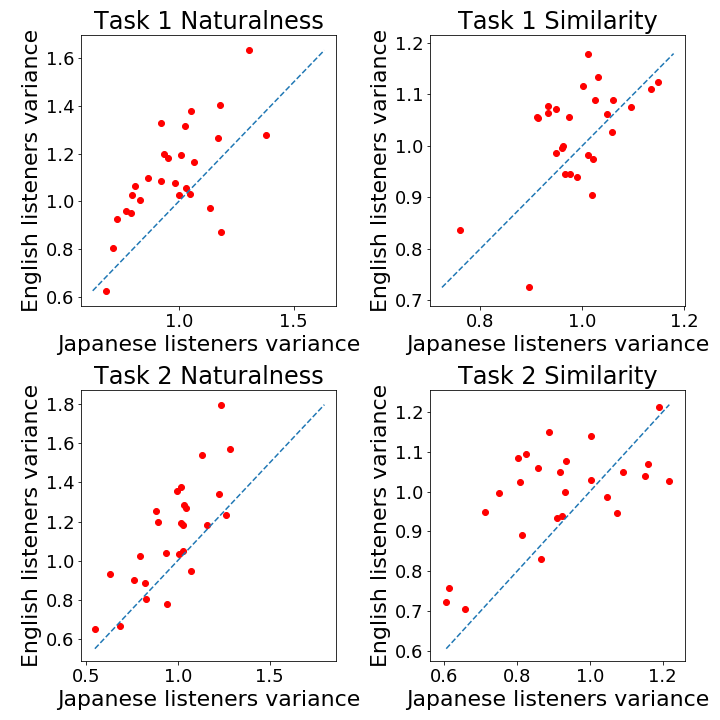}
	\caption{\label{fig:var_enjp}Scatter plots of system-level variance from Japanese and English listeners.
 }	
\end{figure}

\section{Details of the B01 DiffSVC baseline system}
\label{appen:diffsvc}

The first baseline system is much similar to the system presented in the DiffSVC paper \cite{liu2021diffsvc}.
We use a different PPG model, which is a Conformer-based phoneme recognizer containing 7 conformer blocks. The encoder dimension is 256. In total, the PPG model contains 31.2 million trainable parameters. The training data is a combination of a random half from the WenetSpeech dataset (Mandarin Chinese) \footnote{\url{https://wenet.org.cn/WenetSpeech}} and a random half from the GigaSpeech dataset (English) \cite{gigaspeech}, which in total has 10k hours speech data. We take the feature from the last hidden layer as the content feature.

The PPG-to-Mel-spectrogram model has the same network structure as that presented in \cite{liu2021diffsvc}. We extend the model to support multi-singer generation by adding a speaker/singer embedding vector to every residual block.
The training set is a mixture of the SVCC 2023 dataset,  OpenCPOP dataset \cite{opencpop}, MultiSinger \cite{multisinger}, VCTK \cite{vctk}, NUS-48N \cite{nus48e} and M4Singer \cite{m4singer}. In total, the training set contains 116 hours of speech and singing data from 221 speakers or singers. We do not conduct any finetuning procedure for the target singers and use the multi-singer model directly for evaluation. We use a HiFi-GAN V1 \cite{hifigan} to convert the generated Mel spectrogram to a waveform, which is trained with the same training set.

During conversion, for the task of in-domain SVC (i.e., Task 1), we shift the source pitch by multiplying a ratio, which is computed as the ratio of the median of the target pitch and the median of a source phrase. For the task of cross-domain SVC (i.e., Task 2), we shift the source pitch down by an octave in female-to-male conversion and shift the source pitch up by an octave in male-to-female conversion, respectively.

\section{Details of the B02 decomposed FastSVC baseline system}
\label{appen:B02}
For the acoustic model, we use Tacotron 2 \cite{taco2} encoder, along with an autoregressive decoder due to its success in \cite{vcc2020}. The Tacotron 2 encoder consists of two stacks of one-dimension convolutional layers and a bidirectional long short-term memory (BLSTM) layer. On the other hand, the decoder is an autoregressive network, due to its proven ability in the previous challenge \cite{vcc2020}. To implement the autoregressive loop, the previous output is consumed by the first long short-term memory with projection (LSTMP) layer at each time step. The acoustic model predicts the concatenated mel-cepstral coefficients (mcep) and band-aperiodicity (bap), which are used as inputs of the hn-uSFGAN vocoder. For the vocoder, we use HN-uSFGAN \cite{hn-usfgan} as is due to its ability to synthesize waveforms outside the training pitch range.

The network is trained with the SVCC 2023 dataset, along with the large-scale speech dataset VCTK \cite{vctk}, and large-scale singing datasets M4Singer \cite{m4singer}, MultiSinger \cite{multisinger}, OpenCPOP \cite{opencpop}, and NUS-48E \cite{nus48e}. To handle the multilingual datasets, we replace the PPG encoder with HuBERT soft features due to its proven ability in cross-lingual VC \cite{soft-vc-2022}. To optimize the acoustic model, we use two loss functions: 1) an L2 reconstruction loss and 2) a sub-frequency discriminator, which was introduced in \cite{hifisinger}, to improve the predicted mcep/bap features. To shift the pitch, we use linear transformation by using the mean-variance transformation.

\section{Breakdown using different techniques}
\label{appen:breakdown}

Although we mentioned in Sec.~\ref{sssec:taxonomy} that the goal of the taxonomy analysis is not to derive meaningful tendencies or scientific differences, we still tried to find certain techniques that contribute to a high performance. In light of this, we created variants of the scatter plot in Figure~\ref{fig:scatter} by coloring each system with the technique used in that system. The results are shown in Figure~\ref{fig:breakdown}.

We did not find particular trends for the content feature, vocoder, and the use of \textit{so-vits-svc}. We would like to emphasize that, despite the seeming success of \textit{so-vits-svc}, SVC systems based on that toolkit did not necessarily perform better. On the other hand, many of the VAE-based systems had high rankings in Task 1, which somehow shows that VAE can be a promising framework for SVC.

\begin{figure*}[h]
    \centering
    
    \begin{subfigure}[b]{\columnwidth}
        \centering
        \includegraphics[width=\textwidth]{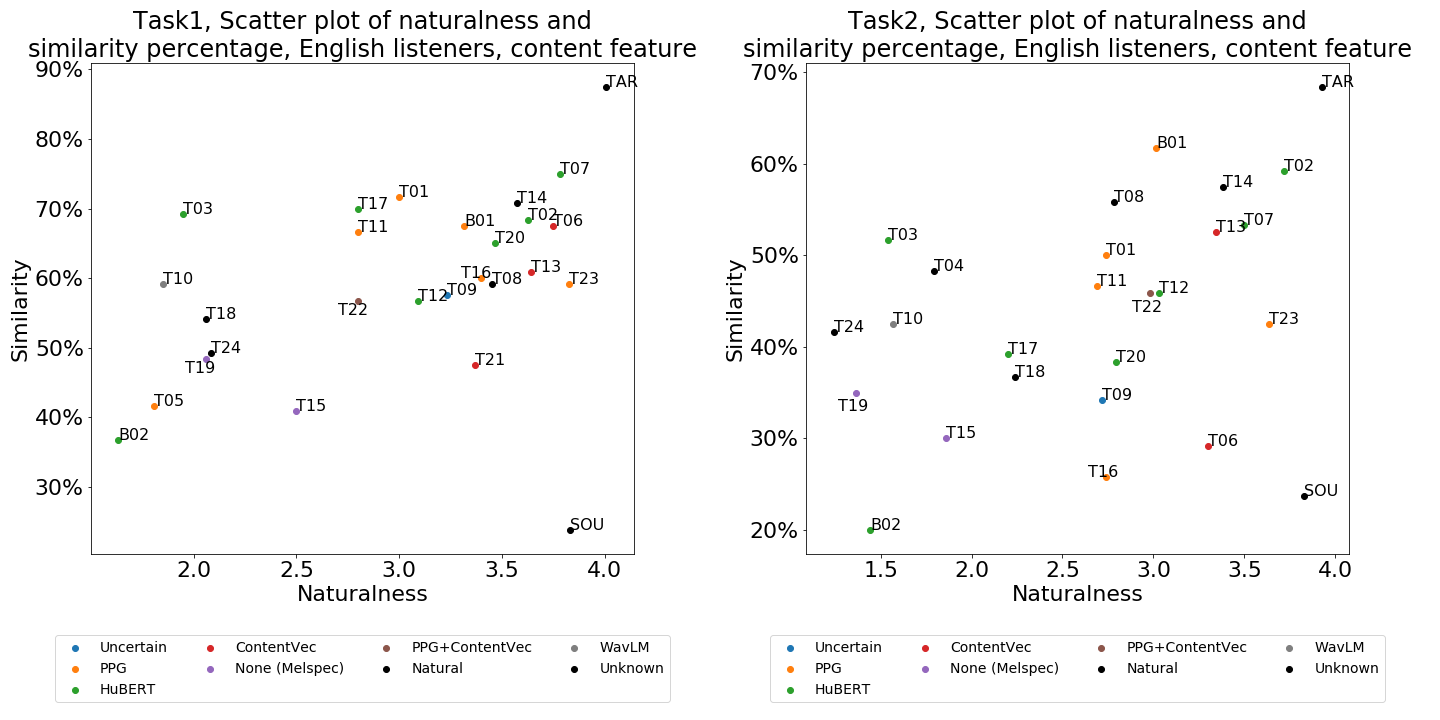}
        \caption{Content feature}
        \label{fig:breakdown_content_feature}
    \end{subfigure}
    \begin{subfigure}[b]{\columnwidth}
        \centering
        \includegraphics[width=\textwidth]{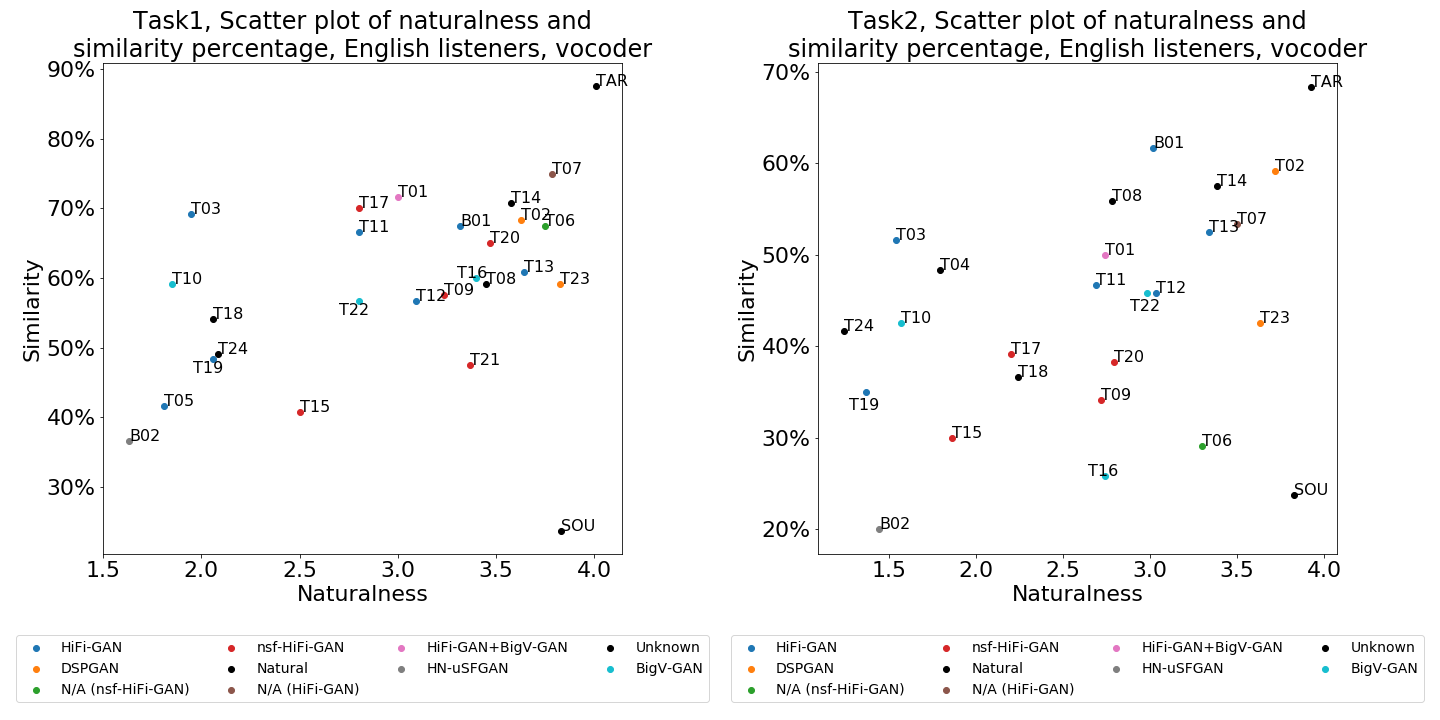}
        \caption{Vocoder}
        \label{fig:breakdown_vocoder}
    \end{subfigure}\\

    \begin{subfigure}[b]{\columnwidth}
        \centering
        \includegraphics[width=\textwidth]{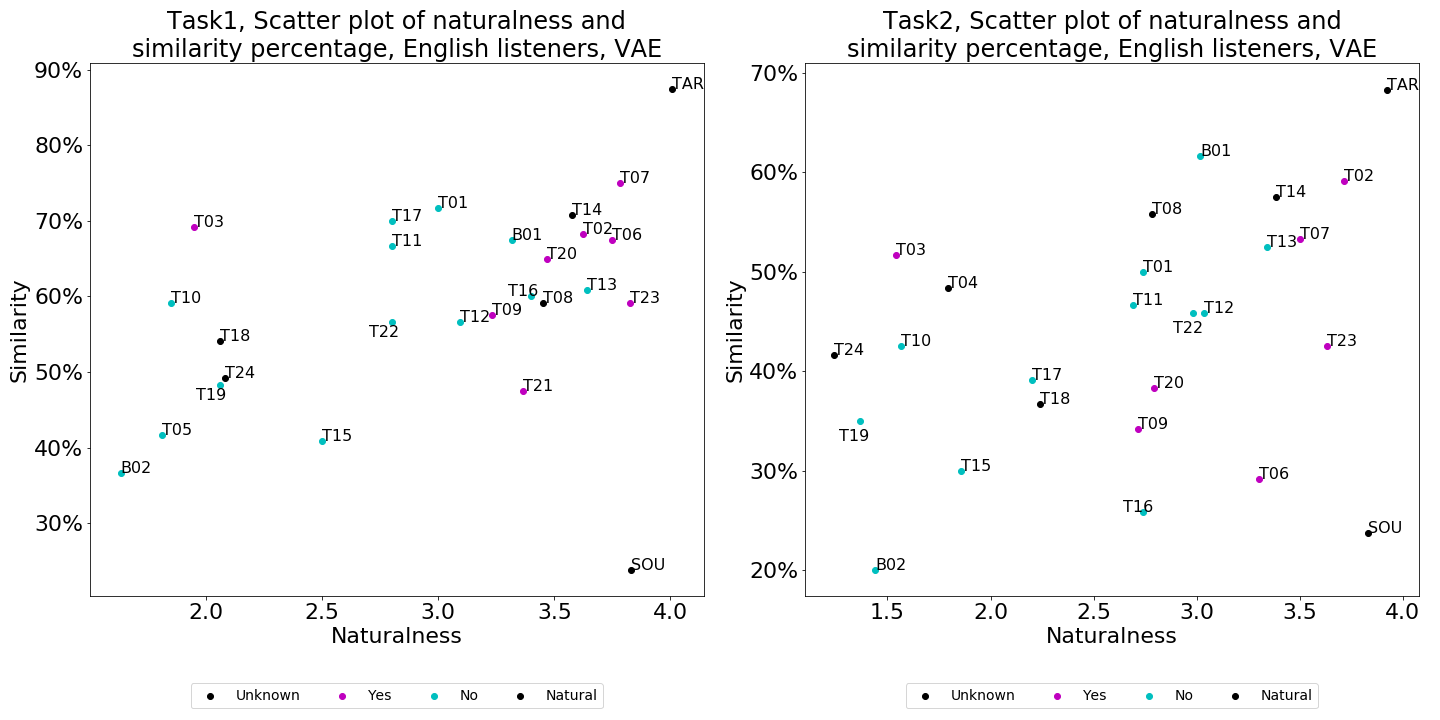}
        \caption{Using VAE or not}
        \label{fig:breakdown_vae}
    \end{subfigure}
    \begin{subfigure}[b]{\columnwidth}
        \centering
        \includegraphics[width=\textwidth]{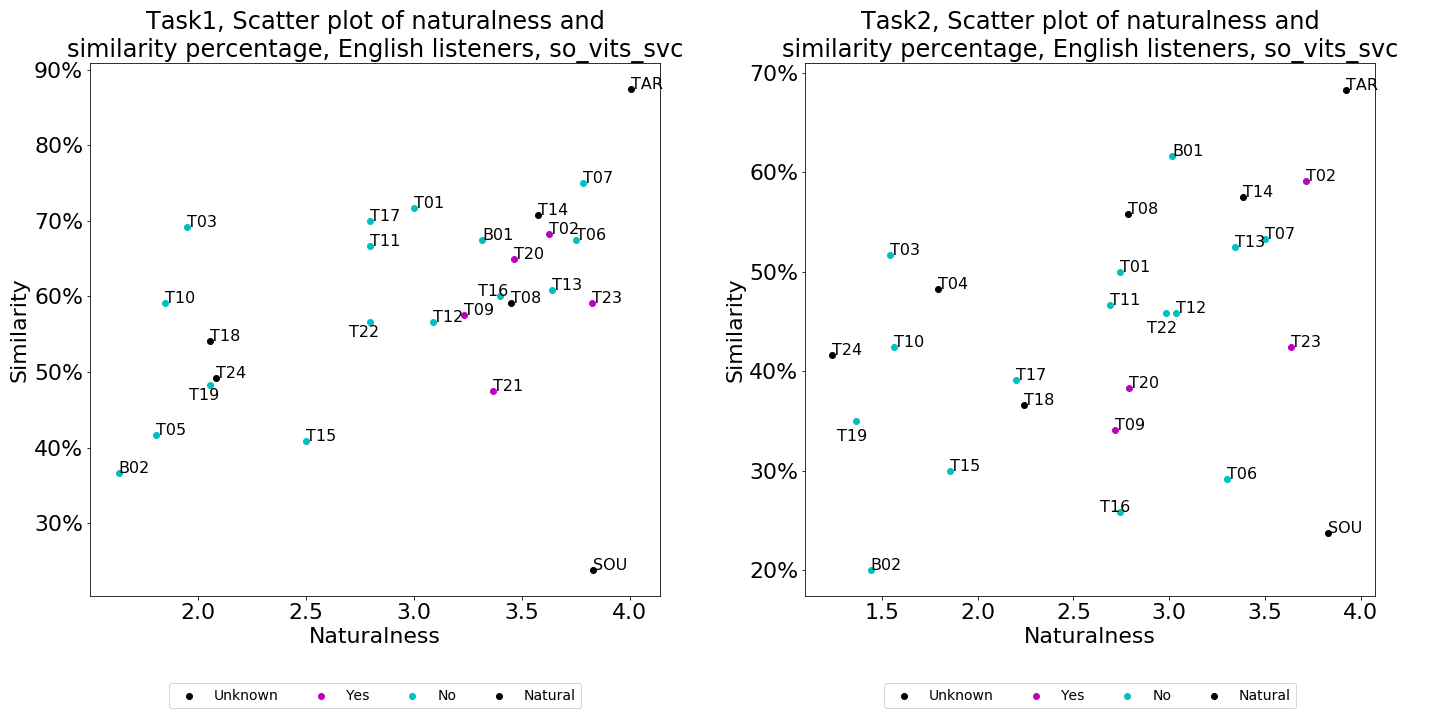}
        \caption{Based on \textit{so-vits-svc} or not}
        \label{fig:breakdown_so_vits_svc}
    \end{subfigure}\\

    \centering
    \caption{Scatter plots of naturalness and similarity in tasks 1 and 2 from English listeners, colored on basis of different techniques.}
    \label{fig:breakdown}
\end{figure*}

\section{Detailed objective evaluation results and analysis}
\label{appen:obj}

Detailed results of the objective evaluation for each team can be found in Table~\ref{tab: task1-detailed-obj} and Table~\ref{tab: task2-detailed-obj}. In general, it is challenging to identify a universally accepted objective measure that correlates strongly with subjective evaluation. This observation is consistent with the findings presented in Table~\ref{tab: spearman} and Table~\ref{tab: pearson}. To further assess performance, we conducted linear regression modeling to examine the impact of different objective measures.

For Task 1, we excluded variables that exhibited high collinearity based on variance inflation factors (VIF). However, for Task 2, we included all variables since no strong collinearity was observed among the different factors. Nonetheless, as shown in Table~\ref{tab: task 1-lr} and Table~\ref{tab: task 2-lr}, although the R-squared values are relatively high, there is limited consensus across different metrics and listening types. One possible reason for the unsuccessful regression could be the limited sample size used in the analysis. But at the same time, these results emphasize the need for further research to identify effective objective evaluation metrics for the challenging SVC task.

\begin{table*}[h]
\centering
\caption{\label{tab: task1-detailed-obj}Detailed objective evaluation results for each team in Task 1.}
\resizebox {\linewidth} {!} {
\begin{tabular}{c|ccccccccc}
\toprule
\textbf{Team ID} & \textbf{MCD($\downarrow$)} & \textbf{F0RMSE($\downarrow$)} & \textbf{F0CORR($\uparrow$)} & \textbf{CER(Conformer)($\downarrow$)} & \textbf{CER(Whisper)($\downarrow$)} & \textbf{$D_{\text{Embed}}$($\downarrow$)} & \textbf{UTMOS($\uparrow$)} & \textbf{SSL-MOS($\uparrow$)}   \\
\midrule
B01 & 11.856 & 70.018 & 0.655 & 34.8 & 23.6 & 0.734 & 1.595 & 0.784 & \\
B02 & 10.409 & 59.028 & 0.664 & 31.3 & 18.5 & 0.454 & 2.028 & 1.072 & \\
T01 & 11.880 & 59.165 & 0.633 & \cellcolor{top5}27.1 & 16.2 & \cellcolor{top5}0.355 & 2.399 & 1.162 & \\
T02 &  \cellcolor{top5}8.524 & \cellcolor{top5}56.187 & \cellcolor{top5}0.722 & 29.8 & 16.1 & 0.439 & 2.345 & 1.112 & \\
T03 &  \cellcolor{top5}9.287 & 78.748 & 0.599 & 38.4 & 33.3 & \cellcolor{top5}0.419 & 1.893 & 1.080 & \\
T05 & 15.867 & 66.427 & 0.671 & 53.4 & 62.0 & 0.647 & 1.917 & 1.275 & \\
T06 & 11.206 & \cellcolor{top5}54.839 & 0.676 & 29.0 & 18.1 & \cellcolor{top5}0.379 & \cellcolor{top5}2.526 & \cellcolor{top5}1.173 & \\
T07 &  \cellcolor{top5}9.427 & 60.757 & 0.680 & \cellcolor{top5}27.7 & \cellcolor{top5}15.0 & \cellcolor{top5}0.357 & \cellcolor{top5}2.420 & \cellcolor{top5}1.193 & \\
T08 &  \cellcolor{top5}9.333 & \cellcolor{top5}57.048 & \cellcolor{top5}0.716 & 28.5 & 18.4 & 0.446 & 2.070 & 1.081 & \\
T09 & 12.155 & 73.416 & 0.587 & 35.0 & 21.4 & 0.477 & \cellcolor{top5}2.598 & 1.145 & \\
T10 & 12.307 & 65.586 & 0.679 & 36.0 & 30.4 & 0.560 & 1.573 & 0.976 & \\
T11 &  \cellcolor{top5}9.160 & 67.867 & 0.671 & 28.8 & 19.1 & 0.428 & 2.246 & 1.113 & \\
T12 & 12.622 & 61.391 & 0.667 & 28.5 & 18.3 & 0.494 & 2.101 & 1.149 & \\
T13 & 10.111 & 68.034 & 0.692 & 30.0 & 18.4 & 0.464 & 2.228 & \cellcolor{top5}1.183 & \\
T14 &  9.762 & 75.901 & \cellcolor{top5}0.686 & 28.7 & \cellcolor{top5}15.6 & 0.448 & 2.057 & 0.999 & \\
T15 & 10.941 & 61.391 & 0.601 & 34.2 & 25.3 & 0.624 & 1.631 & 0.963 & \\
T16 & 10.299 & \cellcolor{top5}55.229 & \cellcolor{top5}0.707 & \cellcolor{top5}24.7 & \cellcolor{top5}12.6 & 0.537 & 2.657 & \cellcolor{top5}1.201 & \\
T17 &  9.454 & 62.937 & 0.657 & 33.3 & 23.1 & 0.427 & 2.082 & 1.109 & \\
T18 & 14.136 & 75.119 & 0.616 & 30.2 & 18.9 & 0.515 & 2.061 & 1.040 & \\
T19 & 10.606 & 75.214 & 0.673 & 33.2 & 23.8 & 0.544 & 1.758 & 0.979 & \\
T20 & 14.229 & 99.252 & 0.643 & 29.9 & 16.2 & 0.484 & \cellcolor{top5}2.313 & 0.971 & \\
T21 & 12.361 & 98.740 & 0.611 & 31.2 & 16.0 & 0.469 & 2.174 & 1.116 & \\
T22 & 11.324 & 57.878 & \cellcolor{top5}0.692 & \cellcolor{top5}26.8 & \cellcolor{top5}15.4 & 0.523 & 2.037 & 1.048 & \\
T23 & 11.536 & \cellcolor{top5}57.784 & 0.678 & \cellcolor{top5}26.8 & \cellcolor{top5}14.5 & \cellcolor{top5}0.423 & \cellcolor{top5}2.717 & \cellcolor{top5}1.323 & \\
T24 & 11.730 & 94.873 & 0.531 & 29.7 & 19.2 & 0.508 & 1.578 & 0.870 & \\
\bottomrule
\end{tabular}
}
\end{table*}

\begin{table*}[h]
\centering
\caption{\label{tab: task2-detailed-obj}Detailed objective evaluation results for each team in Task 2.}
\resizebox {\linewidth} {!} {

\begin{tabular}{c|ccccccccc}

\toprule
\textbf{Team ID} & \textbf{MCD($\downarrow$)} & \textbf{F0RMSE($\downarrow$)} & \textbf{F0CORR($\uparrow$)} & \textbf{CER(Conformer)($\downarrow$)} & \textbf{CER(Whisper)($\downarrow$)} & \textbf{$D_{\text{Embed}}$($\downarrow$)} & \textbf{UTMOS($\uparrow$)} & \textbf{SSL-MOS($\uparrow$)}   \\
\midrule
B01 & 12.495 & 85.693 & 0.243 & 36.3 & 25.0 & 0.761 & 1.679 & 1.023 & \\
B02 & 11.835 & \cellcolor{top5}51.866 & \cellcolor{top5}0.478 & 33.9 & 22.7 & 0.552 & 1.893 & 1.092 & \\
T01 & 12.218 & \cellcolor{top5}52.841 & \cellcolor{top5}0.391 & \cellcolor{top5}25.9 & 26.7 & \cellcolor{top5}0.501 & 2.468 & 1.405 & \\
T02 & \cellcolor{top5}10.278 & 64.436 & 0.254 & 30.5 & \cellcolor{top5}15.5 & 0.551 & 2.415 & 1.223 & \\
T03 & \cellcolor{top5}10.608 & 64.821 & 0.205 & 39.6 & 30.1 & 0.560 & 1.971 & 1.242 & \\
T04 & 12.317 & 63.237 & 0.302 & 31.3 & 20.2 & 0.630 & 1.811 & 0.979 & \\
T06 & 10.651 & 82.362 & 0.356 & 28.1 & 21.4 & 0.577 & \cellcolor{top5}2.870 & \cellcolor{top5}1.792 & \\
T07 & 11.034 & \cellcolor{top5}58.759 & 0.277 & 27.8 & \cellcolor{top5}14.6 & \cellcolor{top5}0.525 & 2.383 & 1.220 & \\
T08 & \cellcolor{top5}10.188 & 66.720 & 0.355 & 29.3 & 23.4 & 0.592 & 2.199 & 1.379 & \\
T09 & 12.448 & 69.065 & 0.322 & 35.5 & 24.5 & 0.584 & \cellcolor{top5}2.715 & 0.969 & \\
T10 & 14.236 & 63.071 & 0.336 & 36.6 & 28.3 & 0.651 & 1.840 & 1.282 & \\
T11 & \cellcolor{top5}10.642 & \cellcolor{top5}53.160 & 0.373 & 29.6 & 20.1 & \cellcolor{top5}0.544 & 2.159 & 1.093 & \\
T12 & 13.281 & 84.585 & 0.351 & 28.5 & 16.4 & 0.576 & 2.084 & 1.178 & \\
T13 & 11.498 & 66.228 & \cellcolor{top5}0.358 & 32.5 & 21.2 & 0.578 & 2.424 & \cellcolor{top5}1.518 & \\
T14 & 11.863 & 65.375 & 0.248 & \cellcolor{top5}27.6 & \cellcolor{top5}15.2 & \cellcolor{top5}0.508 & 2.076 & 1.041 & \\
T15 & 13.331 & 84.585 & 0.284 & 38.8 & 29.0 & 0.776 & 1.986 & \cellcolor{top5}1.476 & \\
T16 & \cellcolor{top5}10.267 & 78.880 & \cellcolor{top5}0.368 & \cellcolor{top5}25.8 & 15.9 & 0.652 & \cellcolor{top5}2.964 & \cellcolor{top5}1.640 & \\
T17 & 10.654 & 81.299 & 0.281 & 30.2 & 28.4 & 0.617 & 2.540 & \cellcolor{top5}1.645 & \\
T18 & 13.983 & 70.474 & 0.260 & 29.1 & 19.8 & 0.566 & 2.416 & 1.200 & \\
T19 & 13.349 & 69.981 & \cellcolor{top5}0.386 & 36.6 & 31.4 & 0.694 & 1.785 & 1.315 & \\
T20 & 14.213 & 71.723 & 0.288 & 29.0 & 21.8 & 0.594 & \cellcolor{top5}2.576 & 1.152 & \\
T22 & 12.314 & \cellcolor{top5}60.867 & 0.252 & \cellcolor{top5}27.1 & \cellcolor{top5}15.4 & 0.549 & 1.966 & 1.020 & \\
T23 & 12.365 & 63.348 & 0.264 & \cellcolor{top5}27.2 & \cellcolor{top5}15.3 & \cellcolor{top5}0.534 & \cellcolor{top5}2.675 & 1.331 & \\
T24 & 13.236 & 90.349 & -0.013 & 34.9 & 28.0 & 0.585 & 1.457 & 1.037 & \\
\bottomrule
\end{tabular}
}
\end{table*}

\begin{table*}[h]
\centering
\caption{\label{tab: pearson}Pearson correlation between objective and subjective metrics. Red highlights indicate the highest correlation with corresponding subjective metrics among the objective metrics. CER metric refers to Conformer-based speech recognition results, while CER+ refers to Whisper results. Significance levels are shown by *. }
\begin{tabular}{l|l|llllllllll}
\toprule
\textbf{Sub. Score} & \textbf{Listener}  & \textbf{MCD} & \textbf{F0RMSE} & \textbf{F0CORR} & \textbf{CER} & \textbf{CER+} & \textbf{$D_{\text{Embed}}$} & \textbf{UTMOS} & \textbf{SSL-MOS}   \\
\midrule
\multirow{2}{*}{Task 1 MOS} & JPN & -0.33 & -0.23 & 0.41** & -0.57*** & -0.58*** & -0.68*** & \cellcolor{high_cor}0.82*** & 0.58*** \\
 & ENG & -0.28 & -0.15 & 0.39* & -0.55*** & -0.57*** & -0.36* & \cellcolor{high_cor}0.66*** & 0.30 \\
 \midrule
\multirow{2}{*}{Task 1 SIM} & JPN & -0.59*** & -0.19 & 0.27 & -0.51** & -0.47** & \cellcolor{high_cor}-0.89*** & 0.51** & 0.35* \\
 & ENG  & -0.43** & -0.16 & 0.27 & -0.41** & -0.40** & \cellcolor{high_cor}-0.46** & 0.37* & 0.04 \\
  \midrule
\multirow{2}{*}{Task 2 MOS} & JPN & -0.39* & -0.37* & 0.37* & -0.71*** & \cellcolor{high_cor}-0.77*** & -0.64*** & 0.69*** & 0.18 \\
 & ENG  & -0.37* & -0.10 & 0.10 & -0.65*** & \cellcolor{high_cor}-0.75*** & -0.38* & 0.59*** & 0.14 \\
  \midrule
\multirow{2}{*}{Task 2 SIM} & JPN & -0.30 & -0.69*** & 0.17 & -0.35* & -0.53*** & \cellcolor{high_cor}-0.71*** & -0.06 & -0.36* \\
 & ENG  & -0.23 & -0.20 & -0.26 & -0.14 & -0.25 & \cellcolor{high_cor}-0.27 & -0.19 & -0.32 \\
\midrule
\multicolumn{9}{r}{Significance levels: ***$p<$0.01, **$p<$0.05, *$p<$0.1} \\
\bottomrule
\end{tabular}
\end{table*}
\begin{table}[h]
\centering

\caption{\label{tab: task 1-lr}Task 1 linear regression models over subjective metrics with objective metrics as inputs. Highlights in orange are coefficients with statistical significance. CER metric refers to Conformer-based speech recognition results.}
\resizebox {\linewidth} {!} {
\begin{tabular}{l|ll|ll}
\toprule
\textbf{Sub. Score} & \multicolumn{2}{c|}{\textbf{Task 1 MOS}} &  \multicolumn{2}{c}{\textbf{Task 1 SIM}} \\
\textbf{Listener} & \textbf{JPN} & \textbf{ENG}  & \textbf{JPN} & \textbf{ENG} \\
\midrule
Intercept & \cellcolor{sig}\hphantom{-}3.211*** & -3.227 &  -1.143 & \hphantom{-}2.478* \\
MCD & -0.266  & -0.094 &  -0.024  & -0.034\\
F0RMSE & \hphantom{-}0.002  & \hphantom{-}0.016 &  \hphantom{-}0.010 & \hphantom{-}0.002\\
F0CORR & \hphantom{-}1.196  & \hphantom{-}4.122 &  \hphantom{-}4.435* & \hphantom{-}1.004 \\
CER & -1.757e-4  & -0.037 &  -0.021 & -0.005\\
$D_{\text{Embed}}$  & \cellcolor{sig}-2.422*** & \hphantom{-}2.637  & -1.758 & -0.540 \\
UTMOS & -0.054  & \cellcolor{sig}\hphantom{-}1.531*** &  \cellcolor{sig}\hphantom{-}1.241*** & \hphantom{-}0.104 \\
\midrule
$R^2$ & \hphantom{-}0.855  & \hphantom{-}0.634 &  \hphantom{-}0.794 & \hphantom{-}0.332 \\
Adjust $R^2$ & \hphantom{-}0.807  & \hphantom{-}0.511 & \hphantom{-}0.726 & \hphantom{-}0.109 \\
F Significance & \hphantom{-}$<$1e-3  & \hphantom{-}0.003 &  \hphantom{-}$<$1e-3  & \hphantom{-}0.237 \\
\midrule
\multicolumn{5}{r}{Significance Levels: ***$p<$0.01, **$p<$0.05, *$p<$0.1} \\
\bottomrule
\end{tabular}
}
\end{table}

\begin{table}[h]
\centering

\caption{\label{tab: task 2-lr}Task 2 linear regression models over subjective metrics with objective metrics as inputs. Highlights in orange are coefficients with statistical significance. CER metric refers to Conformer-based speech recognition results, while CER+ refers to Whisper results.}
\resizebox {\linewidth} {!} {
\begin{tabular}{l|ll|ll}
\toprule
\textbf{Sub. Score} & \multicolumn{2}{|c}{\textbf{Task 1 MOS}} &  \multicolumn{2}{|c}{\textbf{Task 1 SIM}} \\
\textbf{Listener} & \textbf{JPN} & \textbf{ENG}  & \textbf{JPN} & \textbf{ENG} \\
\midrule
Intercept & \hphantom{-}3.376** & \hphantom{-}3.702 &  \cellcolor{sig}\hphantom{-}4.930*** & \hphantom{-}4.771*** \\
MCD & \hphantom{-}0.138 & -0.071 &  -0.033 & -0.068 \\
F0RMSE & \hphantom{-}0.006 & \hphantom{-}0.002 &  -0.006 & -0.005 \\
F0CORR & \hphantom{-}2.392 & -0.571 &  \hphantom{-}0.652 & -0.884 \\
CER & \hphantom{-}0.049 & -0.006 &  \hphantom{-}0.018 & -0.014\\
CER+ & \cellcolor{sig}-0.099*** & -0.085** &  -0.030** & -0.002 \\
$D_{\text{Embed}}$ & -5.179** & \hphantom{-}0.690 &  \cellcolor{sig}-2.517*** & \hphantom{-}0.230 \\
UTMOS & \hphantom{-}0.609* & \hphantom{-}0.683 &  -0.363** & -0.154 \\
SSL-MOS & \hphantom{-}0.501 & -0.068 &  \hphantom{-}0.086 & -0.235 \\
\midrule
$R^2$ & \hphantom{-}0.887 & \hphantom{-}0.678 &  \hphantom{-}0.828 & \hphantom{-}0.343\\
Adjust $R^2$ &  \hphantom{-}0.827 & \hphantom{-}0.506 & \hphantom{-}0.736 & -0.007 \\
F Significance & \hphantom{-}$<$1e-3 & \hphantom{-}0.011 &  \hphantom{-}$<$1e-3 & \hphantom{-}0.488 \\
\midrule
\multicolumn{5}{r}{Significance Levels: ***$p<$0.01, **$p<$0.05, *$p<$0.1} \\
\bottomrule
\end{tabular}
}
\end{table}

\end{document}